\documentclass[sigconf]{acmart}

\usepackage{booktabs} 

\copyrightyear{2017}
\acmYear{2017}
\setcopyright{acmcopyright}
\acmConference{CCS '17}{October 30-November 3, 2017}{Dallas, TX, USA}
\acmPrice{15.00}
\acmDOI{10.1145/3133956.3133972}
\acmISBN{978-1-4503-4946-8/17/10}

\usepackage{amssymb}
\usepackage{amsmath}
\usepackage[ruled,linesnumbered]{algorithm2e}

\usepackage{url}
\usepackage{caption}
\usepackage{subcaption}
\usepackage{float}
\usepackage{textcomp}

\DeclareMathOperator*{\argmax}{arg\,max}

\newcommand{\user}{u}
\newcommand{\User}{\mathcal{U}}
\newcommand{\loc}{\ell}

\newcommand{\Loc}{\mathcal{L}}

\newcommand{\uci}{\tau}

\newcommand{\uloc}{\omega}

\newcommand{\biG}{\mathcal{G}}
\newcommand{\Edg}{\mathcal{E}}

\newcommand{\w}[1]{w_{#1}}

\newcommand{\rwtrace}{\phi}
\newcommand{\Rwtrace}{\Phi}
\newcommand{\Nb}[1]{N(#1)}

\newcommand{\feature}[1]{\theta(#1)}
\newcommand{\Together}{\Delta}

\newcommand{\utilityloss}{\phi}
\newcommand{\utility}{\psi}
\newcommand{\Utility}{\Psi}
\newcommand{\obf}{b}
\newcommand{\Visitloc}{A}

\begin{document}
\title{walk2friends: Inferring Social Links from Mobility Profiles}

\author{Michael Backes}
\affiliation{
\institution{CISPA, Saarland University\\
Saarland Informatics Campus}
}
\author{Mathias Humbert}
\affiliation{
\institution{Swiss Data Science Center\\
ETH Zurich and EPFL}
}
\author{Jun Pang}
\affiliation{
\institution{FSTC and SnT\\
University of Luxembourg}
}
\author{Yang Zhang}
\affiliation{
\institution{CISPA, Saarland University\\
Saarland Informatics Campus}
}

\begin{abstract}
The development of positioning technologies
has resulted in an increasing amount of mobility data being available.
While bringing a lot of convenience to people's life,
such availability also raises serious concerns about privacy.
In this paper, we concentrate on one of the most sensitive information that can be inferred from mobility data, namely social relationships.
We propose a novel social relation inference attack that relies on an advanced feature learning technique
to automatically summarize users' mobility features.
Compared to existing approaches,
our attack is able to predict any two individuals' social relation,
and it does not require the adversary to have any prior knowledge on existing social relations.
These advantages significantly increase the applicability of our attack and the scope of the privacy assessment.
Extensive experiments conducted on a large dataset
demonstrate that 
our inference attack 
is effective,
and achieves between 13\% to 20\% improvement over the best state-of-the-art scheme.
We propose three defense mechanisms -- hiding, replacement and generalization -- and evaluate their effectiveness for mitigating the social link privacy risks stemming from mobility data sharing.
Our experimental results show that both hiding and replacement mechanisms outperform generalization. Moreover, hiding and replacement
achieve a comparable trade-off between utility and privacy,
the former preserving better utility and the latter providing better privacy.
\end{abstract}


\maketitle
\section{Introduction}
\label{sec:intro}

With the widespread usage of portable devices, 
mobility data has become available to a plethora of service providers, 
such as telecommunication operators, credit card companies, 
location-based services and online social networks (OSNs). 
While substantially improving mobile users' experience 
and providing them with convenient services, 
e.g., location recommendation,
such availability also raises serious concerns about privacy.
Previous studies have shown that
a user's mobility trace is higly unique~\cite{MHVB13} 
and can be effectively deanonymized~\cite{SH12} with side channel information,
and that a user's location data can unveil 
his personal attributes~\cite{PZ17} and identity~\cite{HSGH13}.

Social relationships represent highly privacy-sensitive information 
that is deeply connected with our social identity~\cite{CBCSHK10,MPS13}.
In practice, online social network users have realized the extent of this threat and increasingly concealed their social relationships. For instance, the percentage of Facebook users in New York hiding their friend lists increased from 17.2\% in 2010 to 56.2\% in 2011~\cite{DJR12}.
However, many individuals do not yet realize that their mobility data can also unveil their social relationships.
Using location data to infer the underlying social relations between mobile users is of particular interest to various adversarial parties getting access to mobility data but not to social relations.
For example, it is now well known that the NSA collects location and travel habit data 
to find unknown associates of targets it already knows about~\cite{NSA}.

Previous works~\cite{CBCSHK10,ZP15,SNM11,PSL13,BHJLHGN13,WLL14} 
have demonstrated that mobility data can indeed serve as a strong predictor  
for inferring social relationships.
However, these studies are all conducted with a data-mining perspective, e.g., for recommending friends to users in OSNs. 
They notably impose several requirements
on the mobility data needed to infer social links, 
which dramatically reduces the scope of their applicability.
For instance, almost all existing effective methods
can only be applied if two individuals share locations in common.
However, from a privacy point of view,
in order to fully assess the extent to which location data can reveal the social relationships of any possible user, no such requirement should be a priori imposed.
Moreover, no mitigation techniques 
have been proposed and evaluated so far for countering potential adversarial social link inference. 
This paper aims at filling these two essential gaps. 
First, the link prediction system must be as generic as possible to be able to evaluate, for \emph{any} possible mobile user, the extent of the privacy risk towards his social links.
Second, it is of utmost importance to design effective defense mechanisms for reducing the inherent risk towards social link privacy in location-data sharing. 

\smallskip
\noindent\textbf{Inference attack.}
Our link inference attack aims at predicting whether any pair of individuals are socially related,
regardless of whether they have shared any common locations before.
The attack relies on constructing an informative mobility profile/features for each user,
and comparing two users' profiles,
with the assumption that the mobility profiles of friends should be more similar than the profiles of strangers.
However, manually constructing mobility features 
normally involves tedious efforts and domain experts' knowledge.
Instead, we rely on an advanced feature learning model (based on neural networks)
to automatically learn each user's mobility features.
The feature learning method we adopt~\cite{MCCD13,MSCCD13}
is able to preserve a user's \emph{mobility neighbors}, 
containing the locations he has visited, 
and the other users who have visited these locations.
This method assumes that a user's mobility neighbors represent his mobility features to a large extent.
After each user's mobility features are learned,
we utilize pairwise similarity measures to compare two users' features and infer if these users are socially related.
As our inference technique is unsupervised, 
the adversary does not need any prior knowledge on existing social relationships, 
which broadens the range of scenarios our attack can cover.

We empirically evaluate our inference attack on a large-scale dataset 
containing millions of location data points, i.e., check-ins, shared by Instagram users.
Compared to well-known mobility datasets 
containing social relationships \cite{CML11, CCLS11}, our dataset notably includes detailed information about each location, such as 
the location category/semantics.
Extensive experimental evaluation shows that our attack 
is effective (with an area under the ROC curve equal to 0.8), 
and achieves between 13\% to 20\% improvement over the previous methods.
We also empirically study the impact on the performance of various parameters involved in our machine-learning model.
Then, we demonstrate that our attack is robust when a user only shares a small number of locations (down to 5 check-ins),
and can even identify relationships between pairs of users that have shared no common location.
Finally, we show that our attack is also effective when the adversary only has access to coarser-grained mobility data.

\smallskip
\noindent\textbf{Countermeasures.}
In order to mitigate the aforementioned privacy risks, we extend and evaluate three defense mechanisms initially proposed by the location privacy community~\cite{STBH11,BHMSH15}: hiding, replacement and generalization.
In particular, for the replacement mechanism,
we rely on the random walk approach proposed in~\cite{MPS13} to find socially close locations to be replaced with. For generalization, we use two levels of generalization for both the semantic and geographical dimensions~\cite{BHMSH15}. For the inference attack carried out with this countermeasure,
we consider an enhanced adversary who is equipped with background knowledge on each location's popularity.
This allows us to evaluate the generalization mechanism under a realistic setting, and thus have a more meaningful privacy assessment.
We evaluate the effectiveness of the three defense mechanisms on our inference attack as well as on the previously proposed inference methods.

To quantify the utility degradation resulting from our mitigation techniques,
we adopt an information-theoretic metric, 
the Jensen-Shannon divergence,
which measures the difference between each user's location distribution in the original and in the obfuscated dataset.
This utility measurement is meaningful since a user's location distribution
is an essential element for building useful applications from mobility data,
such as location recommendation systems.

Our experimental results show that hiding and replacement
achieve equivalent privacy-utility trade-off:
the former preserves better utility
but the latter can reduce the attack's performance to a larger extent.
Furthermore, both hiding and replacement significantly outperform the generalization mechanism.

\smallskip
\noindent\textbf{Contributions.}
In summary, we make the following contributions.
\begin{itemize}
\item We propose a new social relation inference attack
based on mobility data.
The attack relies on a feature learning method 
and is able to predict any two users' social relationship
regardless of whether they have visited common locations.
This allows us to comprehensively evaluate the social link privacy risks stemming from location sharing.
\item Extensive experiments
demonstrate that our attack significantly outperforms state-of-the-art methods,
and that it is robust to different real-world conditions, including a small number of available location data points.
\item We propose the first defense mechanisms
for protecting social link privacy from mobility-based attacks,
and experimentally demonstrate their effectiveness.
\end{itemize}

\smallskip
\noindent\textbf{Organization.}
Section~\ref{sec:model} presents the notations and
the adversary model considered in this paper.
Our inference attack and its evaluation are presented 
in Sections~\ref{sec:attack} and~\ref{sec:evalua}, respectively.
In Section~\ref{sec:defense}, 
we introduce the defense mechanisms and their evaluation.
Section~\ref{sec:related} presents related work.
We conclude the paper in Section~\ref{sec:conclusion}.

\section{Model}
\label{sec:model}

In this section, we introduce the notations used throughout the paper, as well as the adversary model.

\subsection{User Model}

We typically denote a user by $\user \in \User$ and 
a location by $\loc \in \Loc$ with $\User$ and $\Loc$
representing the sets of users and locations, respectively.
Note that each location considered in this paper 
is mapped to a fine-grained point of interest (POI), 
such as MoMA in New York.
A user $\user$ visiting a location $\loc$ is referred to as a check-in
denoted by a tuple $\langle \user, t, \loc \rangle$,
where $t$ is the time when the check-in happens.
We define $\uci(\user, \loc)$
as the set of all the check-ins of $\user$ at $\loc$,
and $\uci(\user)$ as the set of $\user$'s check-ins in the dataset. 
Moreover, $\uloc(\user)$ is used to denote all the locations $\user$ 
has been to.

\subsection{Adversary model}

The adversary's objective is to infer the social links, or relationships, 
between users by merely observing their mobility data. More precisely, he wants to infer whether two individuals are socially related or not, that is, make a binary prediction on the existence of a social link between two users.
Such adversary can typically represent some location-based services, 
such as telecommunication operators, credit card companies and mobile apps on smart phones,
that collect users' data without having access to their social graph. 

It can also model an OSN user who has access to someone's location check-ins 
but not his social link information.
This is possible on Facebook where a user can choose to hide his friends list,
but keep other information, such as location check-ins, public. Our attack could be used by attackers to learn social links in order to further deanonymize users of the social network(s)~\cite{NS09}.
Finally, it can also represent a global intelligence agency
that gets access to mobility patterns of citizens through their mobile phones' metadata~\cite{NSA}. 

\begin{figure*}[t]
\centering
\includegraphics[width=2\columnwidth]{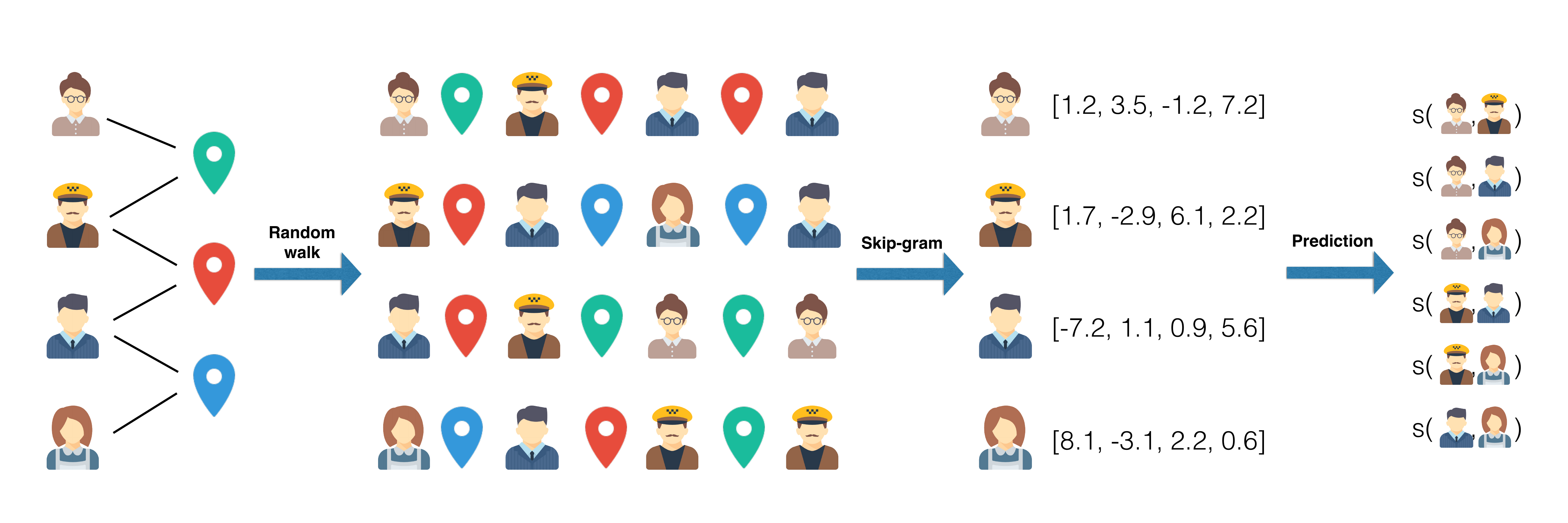}
\caption{Social link inference attack based on location data: a schematic overview.}
\label{fig:attack}
\end{figure*} 

\section{Social Link Inference Attack}
\label{sec:attack}

To infer two users' social relationship with mobility data,
one approach would be to design informative features 
based on the common locations they have visited,\footnote{Two users sharing a common location
indicates that they have both visited the location, regardless of time.}
as proposed in the state-of-the-art works~\cite{SNM11,PSL13,WLL14}.
However, as shown in Section~\ref{sec:evalua}, 
more than 50\% user pairs do not share any common locations, 
meaning that such approaches cannot be applied to infer their social relationships.
Alternatively, we can summarize each user's mobility features (or profile),
then compare two users' features to predict their social link,
with the assumption that friends have more similar mobility profiles than strangers.
This approach enables the adversary to predict any pair of users' social link.
However, defining informative mobility features is a non-trivial task because
it falls into the domain of feature engineering in machine learning,
which normally involves tedious efforts and domain experts' knowledge.
For instance, features such as users' home locations, as proposed in~\cite{SNM11},
have led to poor inference performance (see Section~\ref{sec:evalua}).

The recent advancement of representation/feature learning (deep learning)
provides us with an alternative approach.
In this setting, features are automatically learned following an objective function
that is independent from the downstream prediction task,
in our case, social link inference.
Promising works in this field include~\cite{PAS14,TQWZYM15,GL16},
whose objective functions preserve each user's neighbor information in the social network.
The assumption of these works
is that a user's social neighbors can reflect who he is.
Similarly,
we believe that a user's mobility neighbors
can summarize his mobility profile to a large extent.
Therefore, we utilize feature learning
to automatically learn each user's mobility features,
and apply the learned features for social relation inference.

Our attack can be decomposed into three stages, as depicted schematically in Figure~\ref{fig:attack}.
In the first stage, we adopt a random walk approach 
on the user-location bipartite graph
to obtain random walk traces, which represent each user's neighbors 
in the mobility context.
In the second stage, we feed the obtained random walk traces
to a state-of-the-art feature learning model, namely skip-gram~\cite{MCCD13,MSCCD13},
to obtain each user's mobility features in a continuous vector space.
In the third stage, we measure the pairwise similarity
between two users' vectors to predict whether there exists a social link between them
in an unsupervised setting.

\subsection{Mobility Neighbors with Random Walk}
We organize users and locations
into a weighted bipartite graph $\biG = (\User, \Loc, \Edg)$
where $\Edg\subseteq \User\times\Loc$ contains all the edges between $\User$ and $\Loc$.
For an edge $(\user, \loc)\in \Edg$ between $\user$ and $\loc$,
we define its edge weight $\w{\user, \loc}$
as the number of check-ins of $\user$ at $\loc$, i.e., $\w{\user, \loc}=\vert \uci(\user, \loc)\vert$.
A user's graph neighbors in the mobility context should contain locations he has been to,
especially those locations he frequently visits, but also
indirect neighbors such as other users who have visited the same locations, locations these users have visited, and so on.
It is worth noting that this representation has demonstrated its effectiveness in numerous real-world applications, such as recommendation systems.

To define a user's mobility neighbors,
we could rely on breadth-first sampling (BFS) or depth-first sampling (DFS)~\cite{GL16}.
However, the neighbors resulting from BFS and DFS cannot reflect properly
the user's top visited locations and other users that are similar to him,
as the number of times a user visited a location is not taken into account. 
The random walk method fits our problem better, as it considers edge weights and is computationally more efficient than the aforementioned approaches~\cite{GL16}.
Previously, the random walk approach has been demonstrated to be effective
on homogeneous networks, 
such as social networks, to define a node's neighbors for feature learning~\cite{PAS14,GL16}.
We generalize it to bipartite graphs in this work.

We denote a random walk trace by $\rwtrace$,
which is composed of users and locations
and a set $\Rwtrace$ contains all the random walk traces.
The procedure for generating random walk traces from a user-location bipartite graph 
is listed in Algorithm~\ref{alg:rw}.
For each user, the algorithm generates $t_w$ random walk traces (Line 3),
and each trace is $l_w$ steps long (Line 6).
Here, $t_w$ and $l_w$, referred as \emph{walk times} and \emph{walk length},
are two hyperparameters and their values are set experimentally.
For each current node ${\it curr\_v}$ in a random walk trace,
we extract its neighbors, i.e., ${\it curr\_v\_nb}$, from $\biG$ and 
the corresponding edge weights from ${\it curr\_v}$ to ${\it curr\_v\_nb}$, i.e., ${\it curr\_v\_w}$ (Line 7).
Then, the next node $next\_v$ in the random walk given the current node $curr\_v$ is chosen with the alias method~\cite{W77} according to the following transition probability:
\begin{equation}
P(next\_v\!=\!y\vert curr\_v\!=\!x)\!=\!\begin{cases}
\frac{\w{x, y}}{Z}  &\text{if } x \in \User \wedge (x, y) \in \Edg, \\
\frac{\w{y, x}}{Z} &\text{if } x \in \Loc  \wedge (y, x) \in \Edg, \\
0 &\text{otherwise},
\end{cases}
\end{equation}
where $Z$ is the normalizing constant equal to the sum of the edge weights connected to $x$ (Line 9).
In the end, we obtain $\Rwtrace$ which contains $\vert\User\vert\times t_w$ random walk traces
and each trace is $l_w$ steps long.
The mobility neighbors of a user $\user$, denoted by $\Nb{\user}$,
are the nodes precedent and after $\user$ in all the random walk traces $\Rwtrace$.

\begin{algorithm}[!t]
\KwData{A user-location bipartite graph~$\biG=(\User, \Loc, \Edg)$}
\KwResult{Random walk traces~$\Rwtrace$}
$\Rwtrace\leftarrow [\ ]$\;
\For{$\user\in\User$}{
\For{$i =1$ \KwTo $t_w$}{
$\rwtrace\leftarrow[\user]$\;
${\it curr\_v}\leftarrow \user$\;
\For{$j=2$ \KwTo $l_w$}{
${\it curr\_v\_nb},{\it curr\_v\_w} \leftarrow \mbox{GetNb}({\it curr\_v, \biG)}$;\\
\# extract ${\it curr\_v}$'s neighbors (${\it curr\_v\_nb}$)\\
and the corresponding weights (${\it curr\_v\_w}$)\;
${\it next\_v} \leftarrow \mbox{Sampling}({\it curr\_v\_nb},{\it curr\_v\_w})$\;
append ${\it next\_v}$ to $\rwtrace$\;
${\it curr\_v}\leftarrow {\it next\_v}$\;
}
append $\rwtrace$ to $\Rwtrace$\;
}
}
\caption{Generating random walk traces}
\label{alg:rw}
\end{algorithm}

\subsection{Skip-Gram Model}

In the second stage of our inference attack,
we feed the random walk traces $\Rwtrace$ into the skip-gram model to map each user's mobility information into a continuous vector.
The model outputs one vector per user, which represents his mobility features.
Skip-gram is a (shallow) neural network 
with one hidden layer that preserves a user's graph neighborhood information.
Two users sharing similar mobility neighbors will be closer in the vector space 
(have similar mobility features) under skip-gram,
which makes this model suitable for our prediction task.

The objective function of skip-gram is formalized as follows:
\begin{equation}
\label{equ:Skip-gram}
\argmax_{\theta \in \mathbb{R}^{|\User\cup \Loc | \times d}}\prod_{v\in \User\cup\Loc}{p(\Nb{v}\vert v;\theta)}
\end{equation}
where $\theta$ represents the parameters of the model, 
i.e., the vectors (features) of all nodes in~$\biG$,
and $d$ is the dimension of the learned vectors.
Similar to the walk times $t_w$ and walk length $l_w$ in the first stage,
$d$ is also a hyperparameter that 
we will study in Section~\ref{sec:evalua}.
As we can see from objective function~\ref{equ:Skip-gram},
skip-gram uses each node to predict its neighbor nodes in~$\Rwtrace$.
Next, by assuming that predicting neighbor nodes are independent of each other,
objective~\ref{equ:Skip-gram} can be factorized into:
\begin{equation}
\label{equ:independent}
\argmax_{\theta \in \mathbb{R}^{|\User\cup \Loc | \times d}}\prod_{v\in \User\cup\Loc}\prod_{n\in \Nb{v}}{p(n\vert v;\theta)}.
\end{equation}
The conditional probability $p(n\vert v;\theta)$
is modeled with a softmax function:
\begin{equation}\label{equ:softmax}
p(n\vert v;\theta) = \frac{e^{\feature{n}\cdot \feature{v}}}{ \sum\limits_{m\in \User\cup\Loc} e^{\feature{m}\cdot \feature{v}}}
\end{equation}
where $\feature{v}\in \mathbb{R}^d$ is the vector we aim to obtain for $v$
and $\feature{n}\cdot \feature{v}$ is the dot product of the two vectors.

By plugging softmax into objective function~\ref{equ:independent}
and applying log-likelihood transformation,
skip-gram is turned into:
\begin{equation}
\label{equ:log-like}
\argmax_{\theta \in \mathbb{R}^{|\User\cup \Loc | \times d}}\!\sum_{v\in \User\cup\Loc}\sum_{n\in \Nb{v}}
\left(\!\feature{n}\!\cdot\! \feature{v}\!-\!\log\!\!\!\!\sum_{m\in \User\cup\Loc}\!\!\!\!e^{\feature{m} \cdot \feature{v}}\!\right).
\end{equation}
From objective function~\ref{equ:log-like},
we can observe that if two nodes share similar neighbors,
then their vectors will be similar.
However, due to 
the term $\log\!\sum\limits_{m\in \User\cup\Loc}\!\!\! e^{\feature{m} \cdot \feature{v}}$, solving objective function~\ref{equ:log-like} is computationally expensive
since it requires summation over all nodes in $\biG$.
In order to 
speed up the learning process, we adopt the negative sampling approach~\cite{MSCCD13}.

The negative sampling approach targets a different objective than the original skip-gram model,
which is whether two nodes $n$ and $v$ 
appear together in a random walk trace or not:
$n \in \Nb{v}$ or $n\notin \Nb{v}$.
It is easy to see that this objective can be interpreted as a binary classification,
and we use a random variable $\Together$ to describe the binary choice:
$\Together=1$ if two nodes appear together in any trace in $\Rwtrace$, and $\Together=0$ otherwise.
Then, the new objective function of skip-gram is:
\begin{equation}\label{equ:negative}
\begin{aligned}
\underset{\theta \in \mathbb{R}^{|\User\cup \Loc | \times d}}{\mathrm{argmax}}
\prod_{v \in\User\cup\Loc}\prod_{n \in \Nb{v}}p(\Together=1\ \vert\ n, v;\theta)\cdot \\
\prod_{v \in\User\cup\Loc}\prod_{n \in \Nb{v}'}p(\Together=0\ \vert\ n, v;\theta),
\end{aligned}
\end{equation}
where $\Nb{v}'$ is a sampled set that 
contains nodes which are not the neighbors
of $v$ in $\Rwtrace$.\footnote{We adopt the same method 
as in~\cite{MSCCD13} to sample non-neighbors.}
The conditional probability $p(\Together\ \vert\ n, v; \theta)$
now is modeled as the binary version of softmax, i.e., logistic regression,
which is denoted by:
\begin{equation}\label{equ:logit}
p(\Together\ \vert\ n, v; \theta)=
  \begin{cases}
\frac{1}{1+ e^{-\feature{n}\cdot \feature{v}}} & \mbox{ if } \Together=1,\\
\frac{1}{1+ e^{\feature{n}\cdot \feature{v}}} & \mbox { if } \Together=0.\\
  \end{cases}
\end{equation}
By adding all the pieces together,
we have the following objective function for skip-gram:
\begin{equation}\label{equ:final}
\begin{aligned}
\underset{\theta \in \mathbb{R}^{|\User\cup \Loc | \times d}}{\mathrm{argmax}}
\sum_{v \in\User\cup\Loc}\sum_{n \in \Nb{v}}
\log\frac{1}{1+ e^{-\feature{n}\cdot \feature{v}}} \ +\ \\
\sum_{v \in\User\cup\Loc}\sum_{n \in \Nb{v}'}
\log\frac{1}{1+ e^{\feature{n}\cdot \feature{v}}}.
\end{aligned}
\end{equation}

Compared to objective function~\ref{equ:log-like}, which is a multi-label classification,
objective function~\ref{equ:final} is more efficient to compute.
We apply stochastic gradient descend (SGD) in our experiments to solve it,
which eventually outputs the feature vectors of all the users in the dataset.\footnote{Besides users' vectors,
we also obtain locations' vectors.
As we want to predict users' social links,
the location vectors are simply dropped.}

\subsection{Social Link Prediction}

In the last stage, for each pair of users $u$ and $v$ 
whose social link we aim to predict,
we adopt a pairwise similarity measurement $s$
to compare their feature vectors learned through skip-gram.
We decide that $u$ and $v$ are socially related if their similarity $s(\feature{u}, \feature{v})$ is above a given threshold.
We experimentally compare the effectiveness of various similarity measurements in Section~\ref{sec:evalua}.

To the best of our knowledge,
our attack is the first to utilize pairwise similarity metrics
to infer two users' social relation based on skip-gram learned vectors.
It is also worth noting that the existing feature learning methods~\cite{PAS14,TQWZYM15,GL16}
focus on user-specific prediction tasks, such as user attribute inference,
and rely on supervised learning algorithms.

\subsection{Advantages of Our Approach}
There are three main advantages of our link inference attack.
First, our attack is performed in an unsupervised setting,
i.e., the adversary does not need any prior knowledge about any existing social relationships
among the users.
Second, our method can be applied to predict a social link between any pair of users
without requiring them to share common locations.
Both of these advantages result in our attack being more generic and applicable to large-scale privacy assessment than previous works.
Third, our attack outperforms state-of-the-art attacks significantly, as shown in the next section.

\section{Attack Evaluation}
\label{sec:evalua}

We evaluate our proposed social link inference attack in this section.
We first describe our experimental setup, including dataset, evaluation metric,
baseline models and parameter setting.
Then, we present the general results for the inference,
and experimentally study the sensitivity of the hyperparameters involved in our inference attack.
Next, we evaluate the robustness of our attack 
with respect to the number of check-ins a user shares,
and the number of common locations between two users.
Finally, we assess the performance of our attack when the adversary only has access to coarse-grained location information.

\subsection{Experimental Setup}

\noindent\textbf{Dataset.}
Since we need social relationships to be explicitly disclosed to construct our ground truth,
we rely on OSN data to conduct our evaluation.
Among all the OSNs, we chose Instagram for two reasons.
First, Instagram is the second largest social network with a fast growing number of users,
and its users are more likely to share check-ins than other OSNs'. For instance,
Instagram users share 31 times more their locations than Twitter users~\cite{MHK14}.
Second, Instagram's location service is linked with Foursquare,
a popular location-based social network,
which allows us to collect detailed information about each location such as its name and category.
In particular, the location category information serves as the basis for one of the defense mechanisms,
namely generalization, which will be presented in Section~\ref{sec:defense}.

The data collection was conducted in January 2016.
We concentrate on three major English-speaking cities worldwide: New York, Los Angeles and London.
In the first step, we use Foursquare's API to collect all the Foursquare's location IDs in these cities,
together with these locations' category information.
Then, we use Instagram's API to transform Foursquare's location IDs 
to the corresponding Instagram's location IDs.\footnote{The connection between Instagram's API 
and Foursquare's API was aborted in April 2016 (\url{https://www.instagram.com/developer/changelog/}).}
In the end, we use Instagram's API to extract 
all the users' check-ins at each location in 2015.
In total, 6.3 million check-ins are collected in New York, 4.6 million check-ins in Los Angeles 
and 2.9 million check-ins in London. Furthermore, the dataset includes 35,389 different locations in New York, 31,991 locations in Los Angeles, and 16,802 locations in London.
Each check-in is organized in the following form:
\[
\langle {\it userID}, {\it time}, {\it latitude}, {\it longitude}, {\it locationID}, {\it category}\rangle.
\]

To collect the ground truth, i.e., the social network data,
we utilize Instagram's API to collect all the IDs for the followees of the users 
in the check-in dataset.\footnote{We only collect 
each user's followees not followers for efficiency reasons:
some users in Instagram have millions of followers, such as celebrities,
and Instagram's API only returns 50 followers per request.}
As in many previous works~\cite{KLPM10,CML11,DTWTCRC12},
we consider two users to have a social relation if they mutually follow each other.

Compared to the well-known mobility datasets 
containing explicit social relation information collected from Gowalla~\cite{CML11} and Twitter~\cite{CCLS11},
our dataset has two advantages.
First, our dataset has a denser volume.
We collected more than 13 million check-ins in only three cities,
while the Gowalla dataset contains 6 million 
and the Twitter dataset contains 22 million check-ins in the whole world.
Second, as mentioned above, our dataset contains detailed information about each location,
which both Gowalla and Twitter datasets do not.
For reproducibility purposes, the dataset will be made available upon request.

\begin{figure*}[!t]
\centering
\begin{subfigure}{0.69\columnwidth}
\includegraphics[width=\columnwidth]{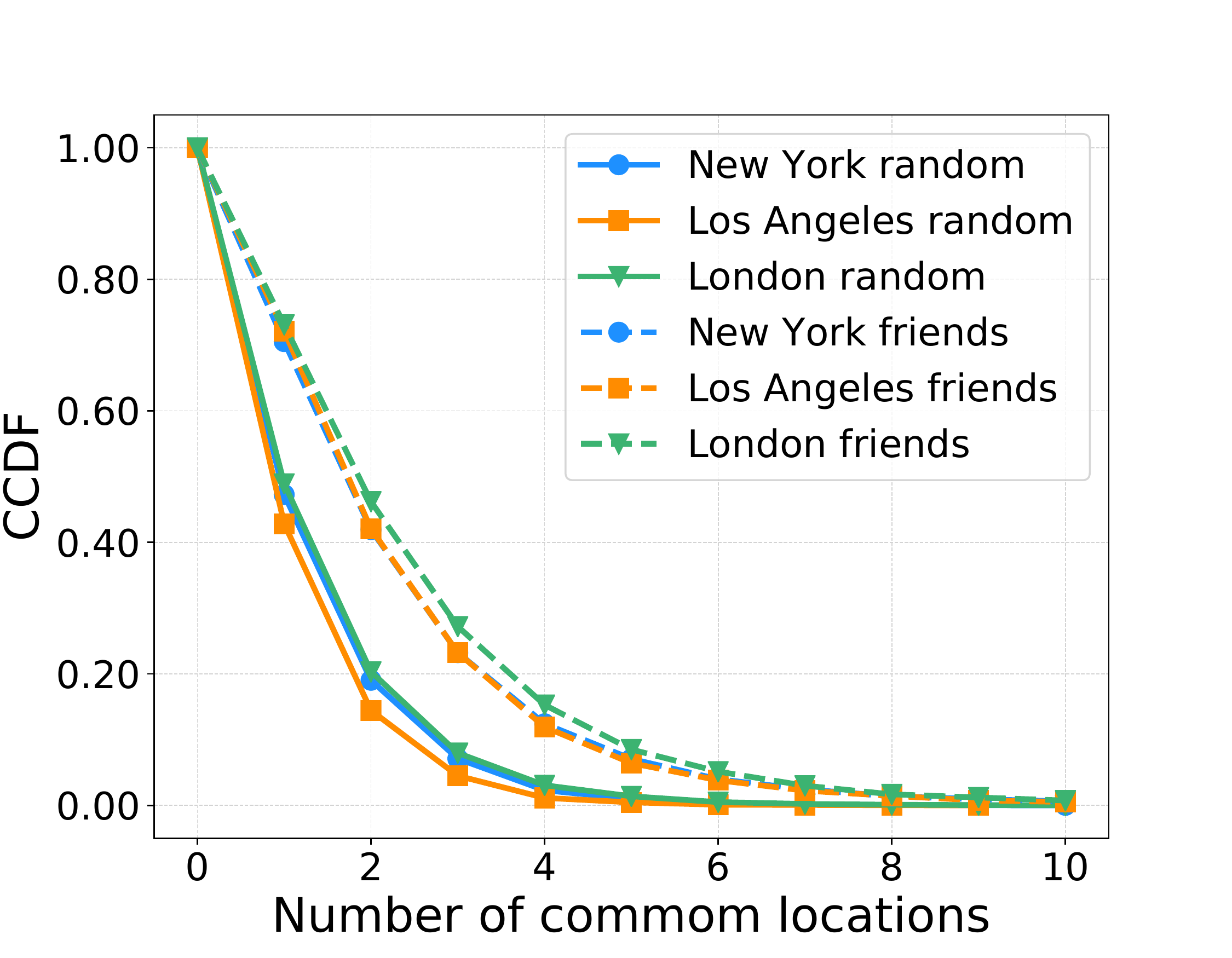}
\caption{}
\label{fig:common_loc}
\end{subfigure}
\begin{subfigure}{0.69\columnwidth}
\includegraphics[width=\columnwidth]{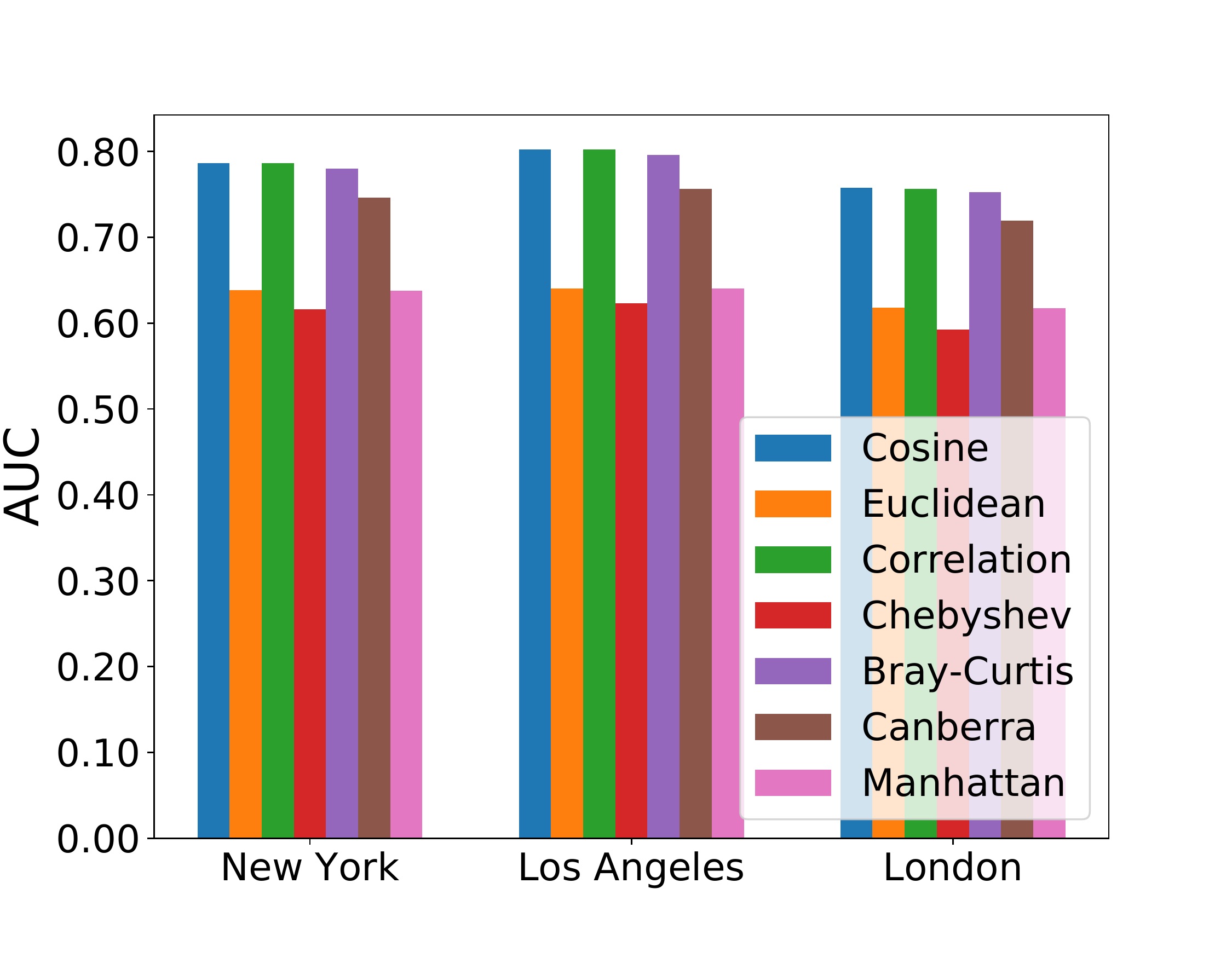}
\caption{}
\label{fig:auc_metric}
\end{subfigure}
\begin{subfigure}{0.69\columnwidth}
\includegraphics[width=\columnwidth]{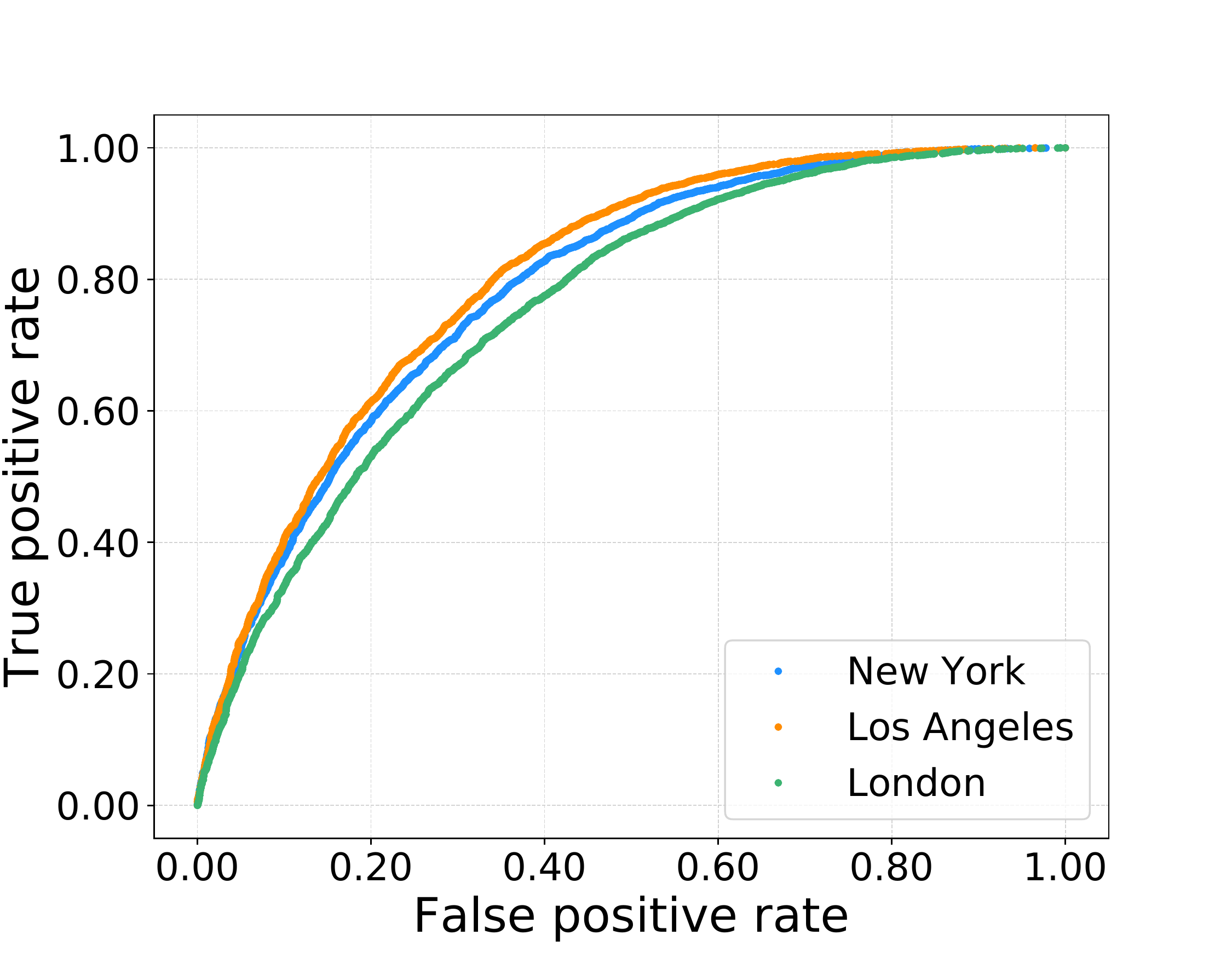}
\caption{}
\label{fig:roc_cosine}
\end{subfigure}
\caption{(a) Distribution of the number of common locations between any two randomly chosen users and two socially related users (friends);
(b) Area under the ROC curve (AUC) with respect to various pairwise similarity measures;
(c) ROC curves with cosine similarity.}
\label{fig:result}
\end{figure*}

\begin{table}[!t]
\centering
\caption{Statistics of the pre-processed dataset.}
\label{table:process}
\begin{tabular}{l | c | c | c}
\hline
& New York & Los Angeles & London\\
\hline
No.\ check-ins & 1,843,187& 1,301,991 & 500,776\\
No.\ locations & 25,868 & 22,260 &10,693\\
No.\ users & 44,371& 30,679 & 13,187\\
No.\ social links & 193,995 &129,004 & 25,413\\
\hline
\end{tabular}
\end{table}

In order to get a representative yet usable dataset, we perform some pre-processing on the collected data.
First, since accounts that share many check-ins at one location 
are generally local businesses, such as restaurants, we filter out users who have not visited at least two \emph{different} locations.
Second, some accounts in Instagram are celebrities or bots who are not the targets of our inference attack,
therefore, we filter out those whose numbers of followers\footnote{
We use Instagram's API to collect each user's number of followers without collecting
the detailed follower list.}
are above the 90th percentile (celebrities)
or below the 10th percentile (bots).
Third, to resolve data sparseness issues, we run most of our experiments on users with at least 20 check-ins, whom we consider to be \emph{active users}.
This is in line with existing works such as~\cite{CCL10,CML11,WLL14,WL16}. However, as there is no standard rule for defining active users ($\geq 20$ check-ins in our case), 
we also study how filtering based on a smaller number of check-ins (down to 5) influences the inference attack's performance 
in Subsection~\ref{subsec:robustness}.
The statistics of our pre-processed dataset is listed in Table~\ref{table:process}.

\smallskip
\noindent\textbf{Metric.}
We adopt AUC (area under the ROC curve) as our attack evaluation metric
for two reasons.
First, due to the nature of social networks,
link inference has a huge prediction space
and the labels are highly imbalanced,
e.g., there are more than 9.8 billion pairs of active users in New York
and less than 0.02\% of them are friends (Table~\ref{table:process}).
To tackle this problem,
we adopt the down-sampling strategy used in~\cite{LLC10,GL16},
that is, we randomly sample the same number of stranger pairs as the number of friend pairs.
To properly evaluate the inference in the down-sampled prediction space,
a metric that is not sensitive to the label distribution is needed.
As pointed out in~\cite{ME11,LLC10}, AUC satisfies this requirement,
and previous inference algorithms~\cite{SNM11,WLL14,GL16} have adopted it for evaluation too.
Second, there exists a conventional standard
for interpreting AUC (whose range is [0.5, 1]): 0.5 is equivalent to random guessing, 1 is perfect guessing (100\% true positives and no false positives), and 0.8 represents already a good prediction.\footnote{\url{http://gim.unmc.edu/dxtests/roc3.htm}} This allows us to intuitively get a sense of the attack's performance, even without comparing against baseline models. Finally, note that privacy is defined as the opposite of the attack success. This means that privacy is minimal when AUC equals 1, and maximal when AUC equals 0.5.

\smallskip
\noindent\textbf{Baseline models.}
We consider 14 baseline models proposed in three state-of-the-art papers inferring
social relationships with mobility data~\cite{SNM11,PSL13,WLL14}.
They are denoted by $\mathtt{common\_p}$~\cite{PSL13}, $\mathtt{overlap\_p}$~\cite{PSL13}, $\mathtt{w\_common\_p}$~\cite{PSL13},
$\mathtt{w\_overlap\_p}$~\cite{PSL13}, $\mathtt{aa\_ent}$~\cite{PSL13}, $\mathtt{min\_ent}$~\cite{PSL13},
$\mathtt{aa\_p}$~\cite{PSL13}, $\mathtt{min\_p}$~\cite{PSL13}, $\mathtt{geodist}$~\cite{PSL13}, $\mathtt{w\_geodist}$~\cite{PSL13}, $\mathtt{pp}$~\cite{PSL13},
$\mathtt{diversity}$~\cite{SNM11}, $\mathtt{w\_frequency}$~\cite{SNM11,WLL14} and $\mathtt{personal}$~\cite{WLL14}.
The formal definitions of these baseline models can be found in their original papers.
Each of the baseline models rely on manually-designed features,
thus can be evaluated in an unsupervised setting as well.

Among all the baseline models,
7 of them ($\mathtt{aa\_ent}$, $\mathtt{min\_ent}$,
$\mathtt{aa\_p}$, $\mathtt{min\_p}$,
$\mathtt{diversity}$, $\mathtt{w\_frequency}$ and $\mathtt{personal}$)
require that two users share at least one common location,
in order to infer whether there is a social link between them or not. 
However, Figure~\ref{fig:common_loc} shows that 
more than half of the active user pairs and around 30\% of friends' pairs  
do not share any common locations in each city.
Therefore, to evaluate these 7 baselines,
we first apply them on pairs of users who share at least one location,
then randomly guess the rest of the pairs' social relationships.\footnote{
The use of random guessing is due to the fact that 
our prediction is conducted in the down-sampled space.}

\smallskip
\noindent\textbf{Parameter settings.}
As presented in Section~\ref{sec:attack},
our model mainly involves three hyperparameters: walk length $l_w$,
walk times $t_w$ and feature vectors' dimension $d$. We set their default values to
$l_w=100$, $t_w=20$ and $d=128$. and evaluate how different values affect the attack performance in Section~\ref{subsec:sensitivity}.
Another parameter is the size of the neighbor nodes in the random walk traces, i.e., $\vert\Nb{v}\vert$.
Following~\cite{PAS14,GL16}, we set it to 20, considering 10 nodes preceding and 10 nodes after $v$ in $\Rwtrace$.
Finally, the learning rate for SGD is set to 0.025.
The source code of our implementation 
is available at~\url{https://github.com/yangzhangalmo/walk2friends}.

\subsection{Social Link Inference}

\begin{figure*}[!t]
\centering
\begin{subfigure}{\columnwidth}
\includegraphics[width=1\columnwidth]{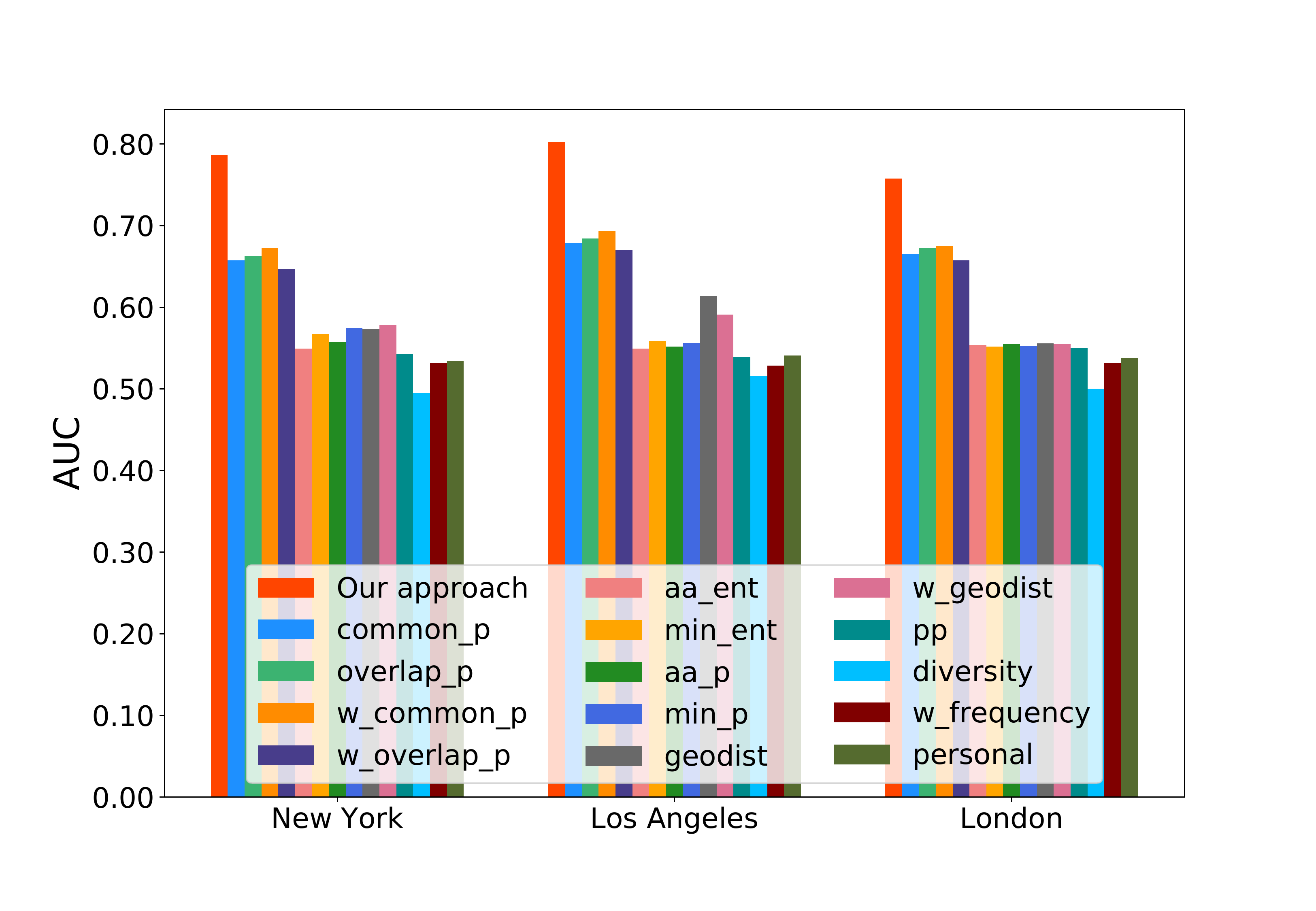}
\caption{}
\label{fig:bl_gen}
\end{subfigure}
\begin{subfigure}{\columnwidth}
\includegraphics[width=1\columnwidth]{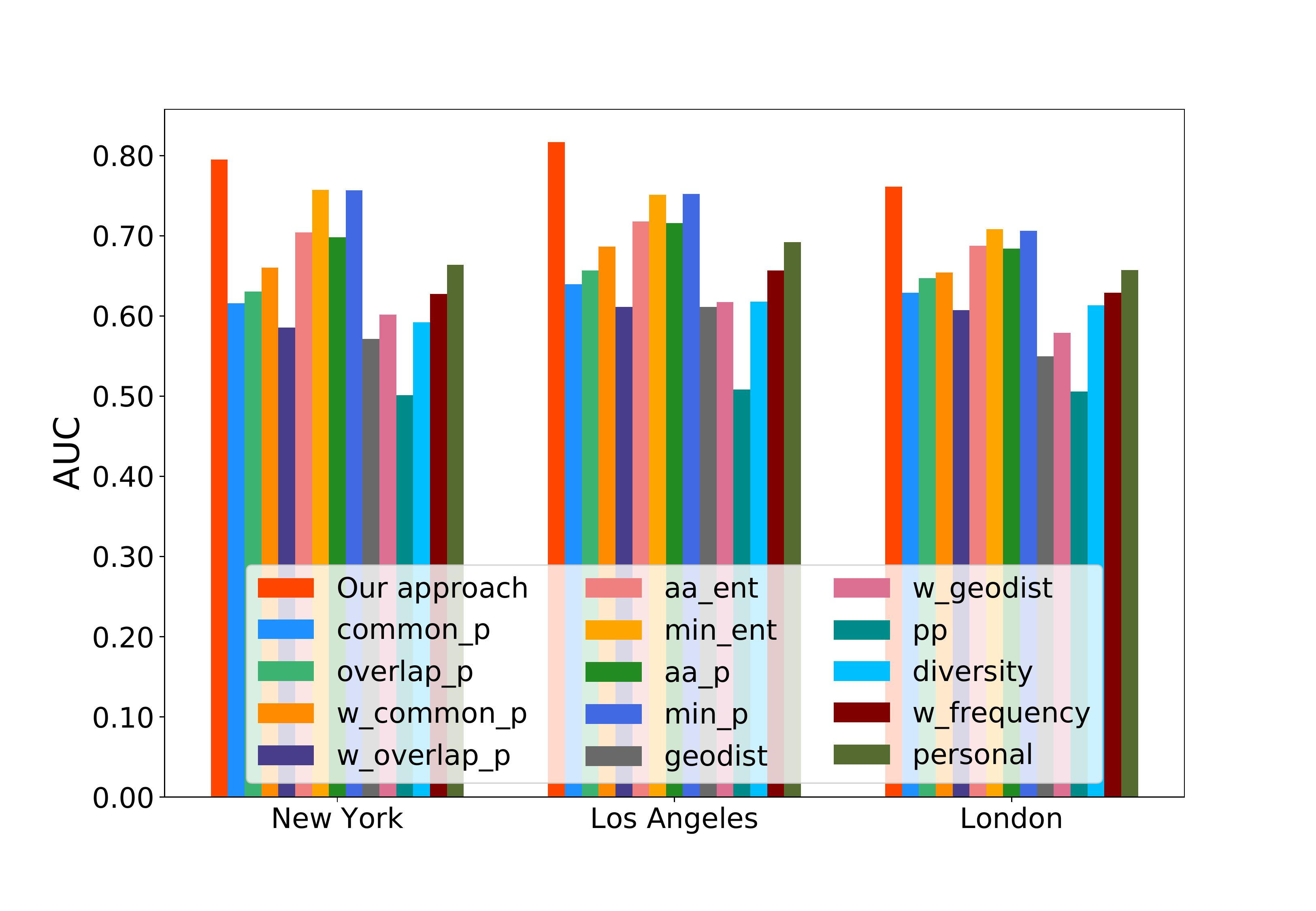}
\caption{}
\label{fig:bl_cp}
\end{subfigure}
\caption{Comparison of our attack against baseline models: (a) using all users,
(b) using only pairs of users who share at least one common location.}
\label{fig:baseline}
\end{figure*} 

\begin{figure*}[!t]
\centering
\begin{subfigure}{0.69\columnwidth}
\centering
\includegraphics[width=\columnwidth]{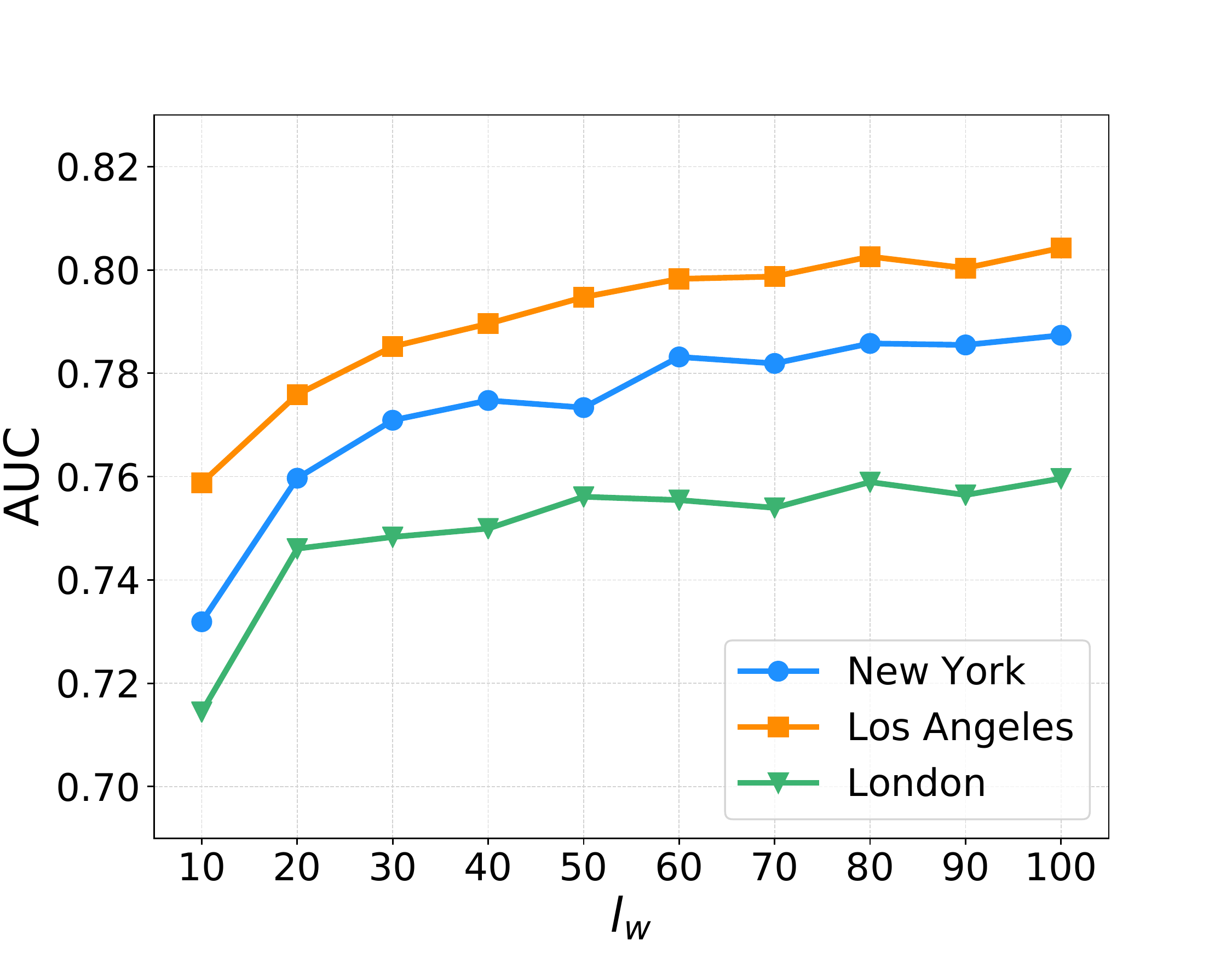}
\caption{}
\label{fig:parameter_walk_len}
\end{subfigure}
\begin{subfigure}{0.69\columnwidth}
\centering
\includegraphics[width=\columnwidth]{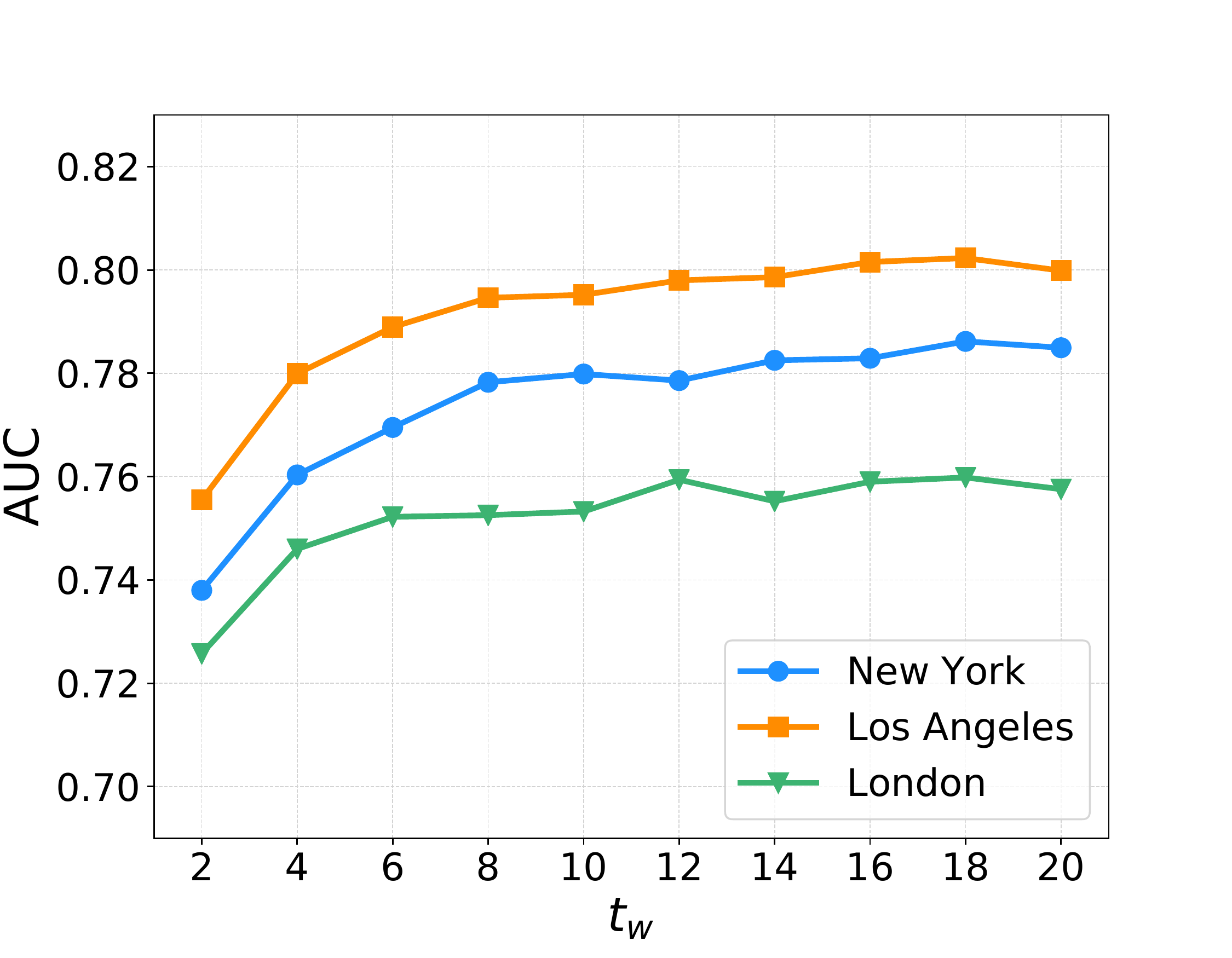}
\caption{}
\label{fig:parameter_walk_times}
\end{subfigure}
\begin{subfigure}{0.69\columnwidth}
\centering
\includegraphics[width=\columnwidth]{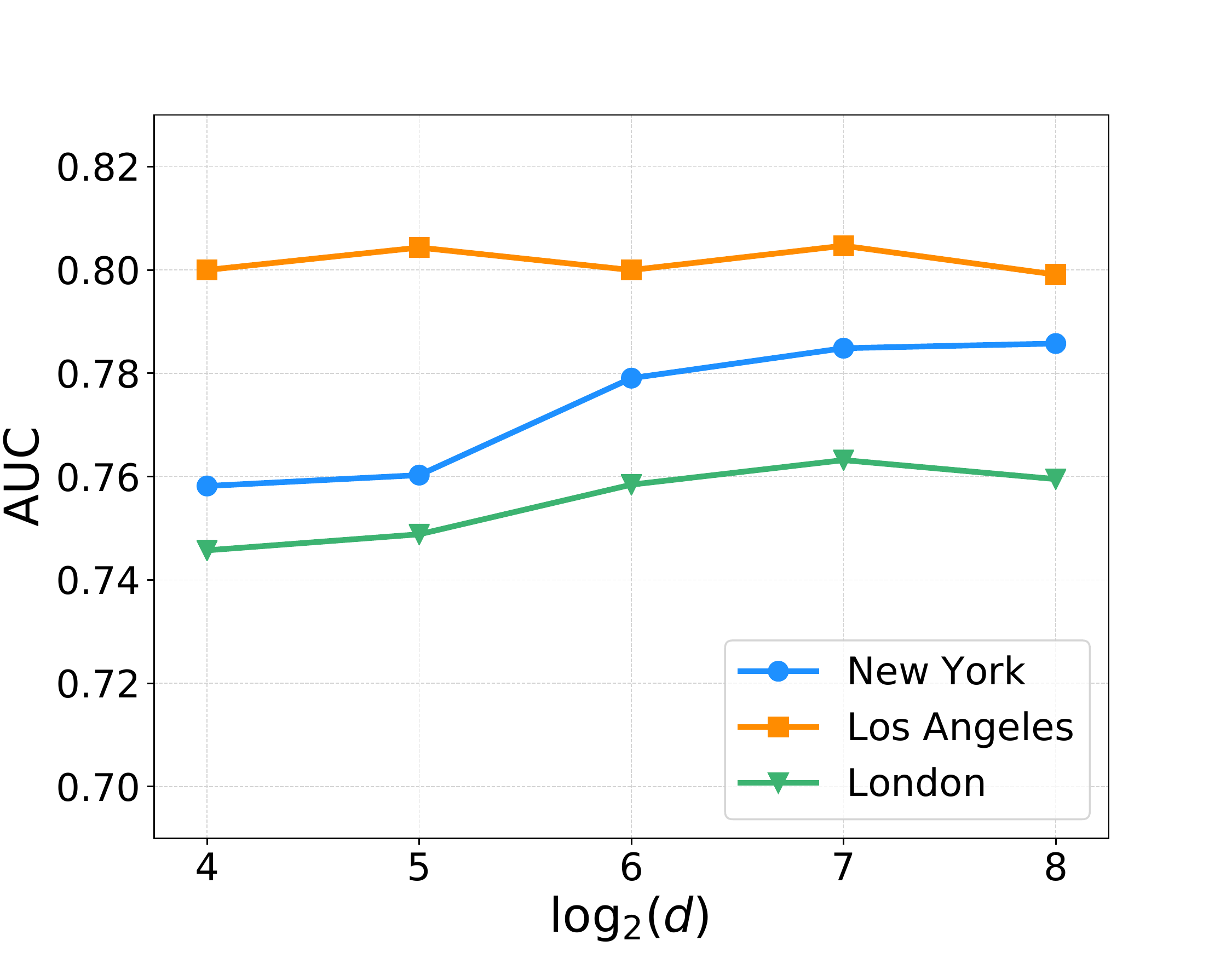}
\caption{}
\label{fig:parameter_num_features}
\end{subfigure}
\caption{Influence of parameters (a) walk length,
(b) walk times and (c) dimension of feature vectors on the inference performance.}
\label{fig:parameter}
\end{figure*} 

Our social link inference attack relies on the pairwise similarity
between two users' mobility features learned by the skip-gram model.
We have evaluated 7 common distance or similarity measures:\footnote{
The formal definitions of these distances are in Appendix~\ref{appendix:pairwise}.}
 cosine similarity, Euclidean distance, correlation coefficient,
Chebyshev distance, Bray-Curtis distance,
Canberra distance, and Manhattan distance.
The corresponding AUC values are depicted in Figure~\ref{fig:auc_metric}.
Among these measures,
cosine similarity, correlation coefficient 
and Bray-Curtis distance achieve the best performance with AUC near 0.8,
which represents a good prediction result.
On the other hand, Chebyshev distance performs the worst with AUC around 0.6.
By looking into all the similarity measures' definition,
we notice that the best performing ones are those 
whose values are bounded.
For instance, correlation coefficient lies within the range [-1, 1].
This indicates that bounded similarity measures provide better results for link prediction based on mobility data.

As cosine similarity maximizes the attack success,
we use it for our inference attack in the rest of this work.
Figure~\ref{fig:roc_cosine} shows the ROC curves 
corresponding to the cosine similarity's AUCs in Figure~\ref{fig:auc_metric}.
The inference performs slightly better for Los Angeles than for New York or London
with the true positive rate being 0.8 while the false positive rate 
staying at 0.34.
The threshold at this point is equal to 0.86, i.e., inferring user pairs
whose features' cosine similarity is above 0.86 as friends leads to a good prediction.

We then compare our inference attack against all the baseline models,
Figure~\ref{fig:bl_gen} shows that
our attack outperforms all the baseline models significantly.
For the best performing baseline model, i.e., $\mathtt{w\_common\_p}$,
we achieve a 20\% performance gain in Los Angeles,
and a 17\% gain in New York.
In the worst case, i.e., London, the performance gain is still 13\%.
This shows that our attack is much more effective than the existing state-of-the-art attacks.

As discussed before,
7 baseline models can only be applied to pairs of users who share common locations.
We further compare our attack against them (as well as the other baselines)
on pairs of users with at least one common location.
Figure~\ref{fig:bl_cp}
shows that these baselines' performances indeed increase
as reported in the original papers,
but our prediction still outperforms the best baseline model, in this case $\mathtt{min\_ent}$,
by 9\% in Los Angeles, 5\% in New York and 7\% in London.
By taking into account the fact that our attack can predict any pair of users' social link,
this further demonstrates the effectiveness of our attack.

\begin{figure*}[!t]
\centering
\begin{subfigure}{0.69\columnwidth}
\centering
\includegraphics[width=\columnwidth]{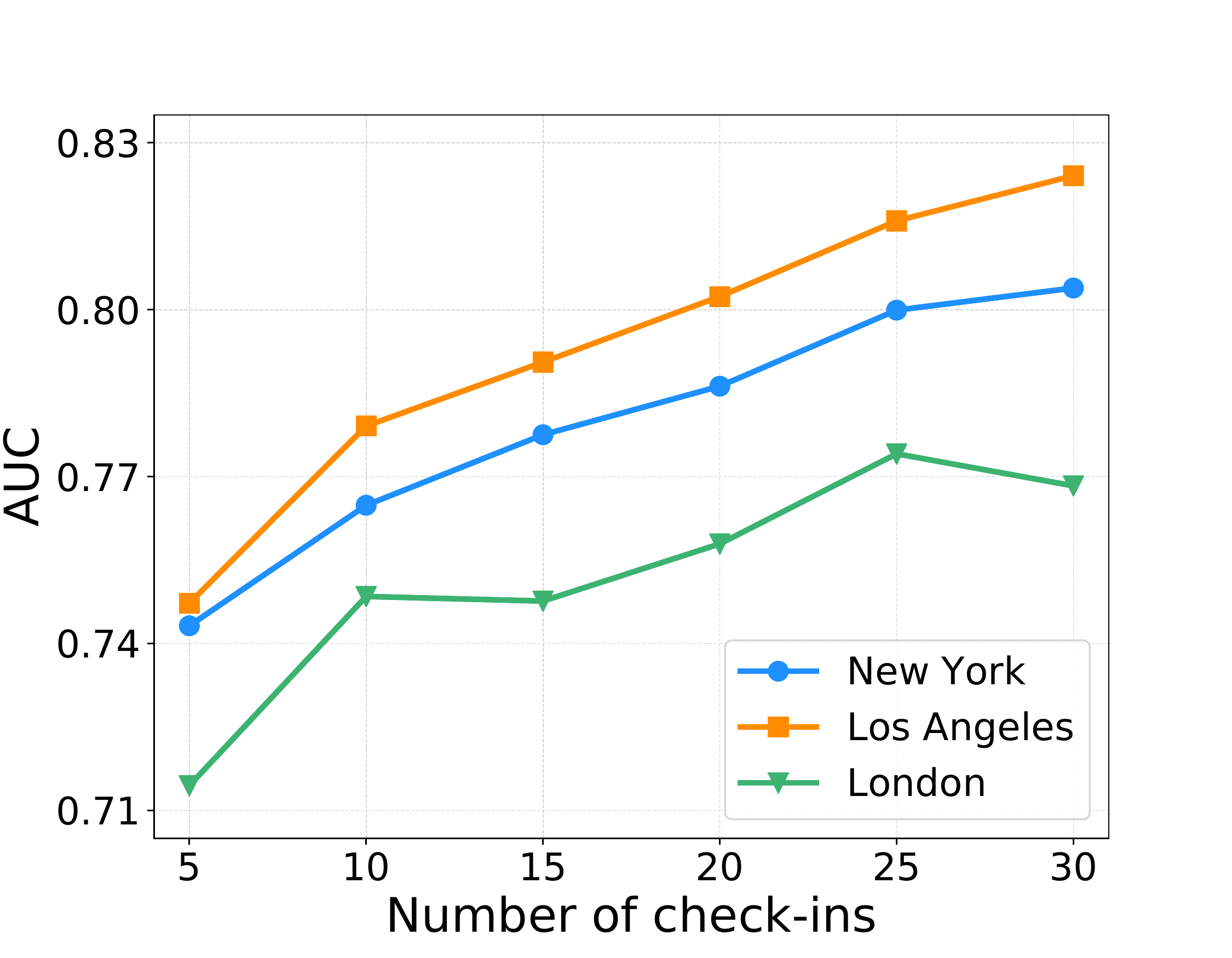}
\caption{}
\label{fig:robust}
\end{subfigure}
\begin{subfigure}{0.69\columnwidth}
\centering
\includegraphics[width=\columnwidth]{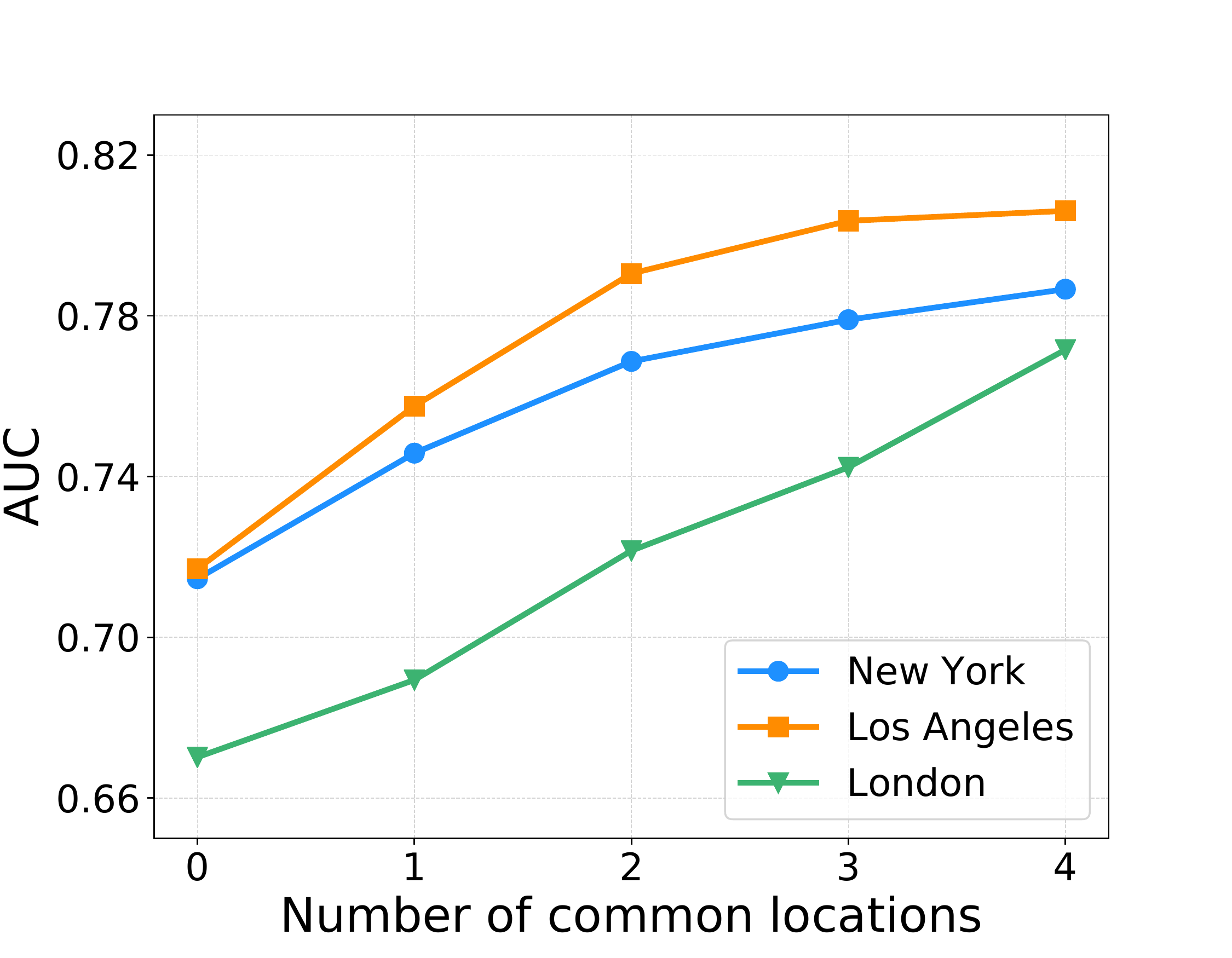}
\caption{}
\label{fig:common_p_auc}
\end{subfigure}
\begin{subfigure}{0.69\columnwidth}
\centering
\includegraphics[width=\columnwidth]{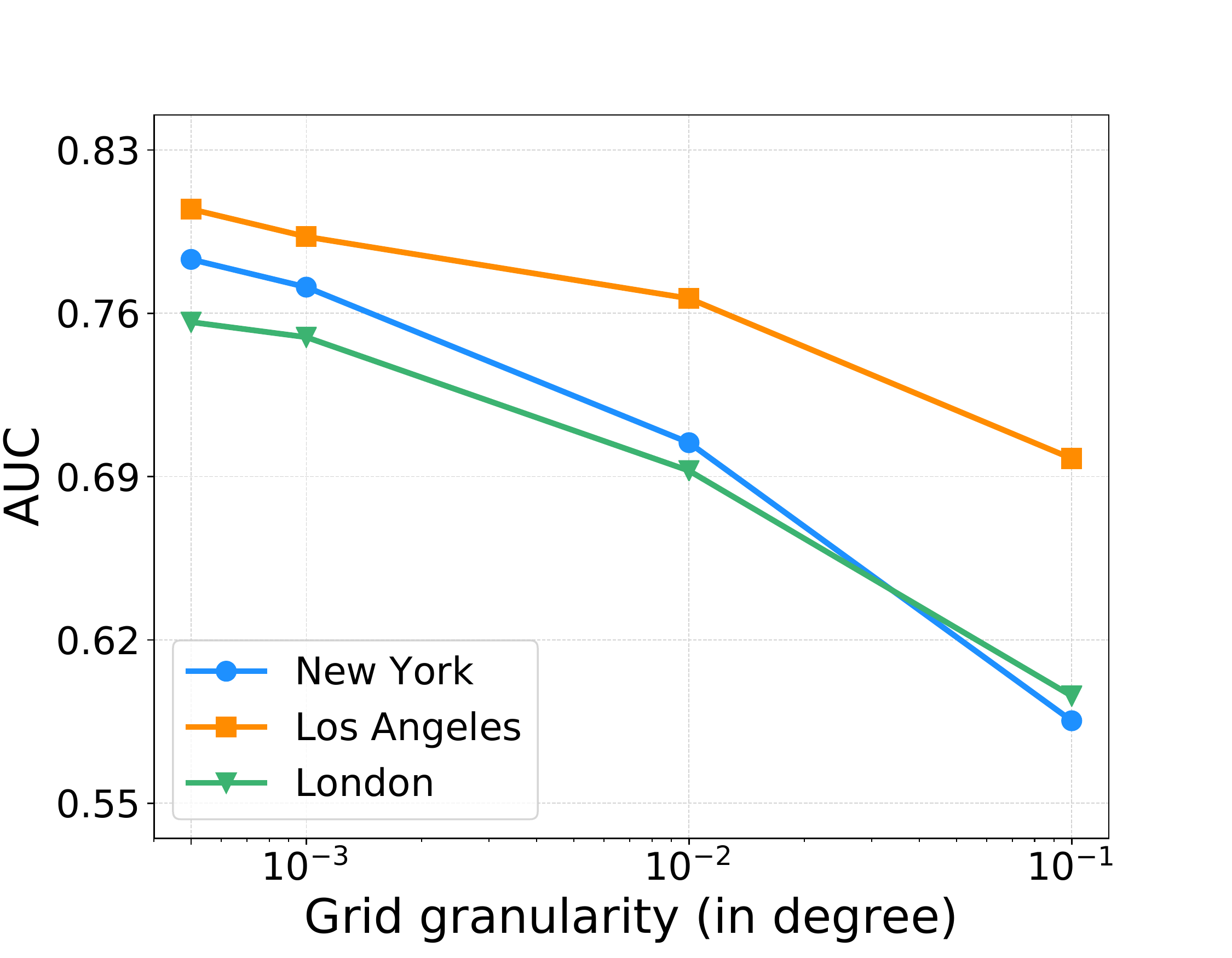}
\caption{}
\label{fig:grid_auc}
\end{subfigure}
\caption{Evolution of the attack performance with respect to 
(a) the minimal number of check-ins shared by every user,
(b) the number of common locations two users share
and (c) different granularities of the geographic grids.}
\end{figure*}

\subsection{Parameter Sensitivity}\label{subsec:sensitivity}
Next, we examine how the different choices of the three hyperparameters 
walk length ($l_w$), walk times ($t_w$) and dimension of feature vectors ($d$) 
affect our attack performance.
When testing each parameter, the two remaining ones are kept to their default settings, 
i.e., $l_w=100$, $t_w=20$ and $d=128$.

Among all the three hyperparameters,
$l_w$ and $t_w$ are directly linked with the size of the random walk traces,
i.e., the amount of data being fed into skip-gram.
Intuitively, larger values of $l_w$ and $t_w$ should lead to better inference performance.
This is indeed the case as shown 
in Figures~\ref{fig:parameter_walk_len} and~\ref{fig:parameter_walk_times}.
The AUC values in all three cities increase sharply 
when $l_w$ increases from 10 to 50, and saturates afterwards.
Similarly, increasing $t_w$ from 2 to 10 leads to around 5\% performance gain in all the cities.

The effect of the vector dimension ($d$), on the other hand, is more subtle.
Previous studies~\cite{MSCCD13,PAS14,GL16} have shown
that larger $d$ results in better performance on node-level prediction.
However, the last stage of our inference attack relies on measuring two vectors' pairwise similarity,
in this case longer vectors do not always yield better performance,
as observed in other data domains such as biomedical data~\cite{BBHHKM16}.
As depicted in Figure~\ref{fig:parameter_num_features},
AUC is rather stable when increasing $d$ compared to $l_w$ and $t_w$,
especially for Los Angeles:
regardless of the choice of $d$, AUC stays around 0.80.
In conclusion, our default hyperparameter settings are suitable
for our inference attack.

\subsection{Attack Robustness}\label{subsec:robustness}

\noindent\textbf{Number of check-ins.}
As discussed above,
our inference attacks are performed on active users,
i.e., users with at least 20 check-ins in each city.
This is in line with the existing works 
on social relation inference attacks and mining user check-ins in general.
However, the optimal definition for active users,
i.e., how many check-ins a user should at least share, is not clear.
The authors of~\cite{CML11} use 10,
\cite{WL16} uses 40, and~\cite{WLL14} uses the top 5,000 users with the most check-ins.
To demonstrate that our attack's performance is robust under all circumstances,
we further study the different choices for defining active users
with respect to AUC.

Figure~\ref{fig:robust} shows that, as we increase the minimum number of check-ins,
AUC increases almost linearly, especially for New York and Los Angeles.
This is expected since the more check-ins a user shares,
the more accurately the adversary can profile his mobility,
which in turn leads to a better social relation inference.
More importantly,
even when concentrating on users with at least 5 check-ins,
our inference attack still achieves a strong performance, e.g., AUC is near 0.75 in Los Angeles.
This indicates that our model can effectively infer a large number of individuals' social links,
and it shows the extent of the privacy threat carried by mobility data at a large scale.
We also discover that the performance differences between our attack
and the best baseline models 
are consistent under different active user definition.
These results demonstrate the robustness of our inference attack.

\smallskip
\noindent\textbf{Number of common locations.}
One of the major advantages of our inference attack
is that it can predict any two individuals' social relationship
regardless of whether they share common locations. Nevertheless, we expect that two users sharing many locations in common will be more likely to be socially related than two users sharing none.
Therefore, we evaluate here how our inference attack performs
with respect to the number of locations users have in common. Recall that, by common location, we mean any location where two users have checked in, not necessarily at the same time.

We select a subset of pairs of users who share between 0 and 4 locations in common, which includes the vast majority of pairs of users (see Figure~\ref{fig:common_loc}), and show the results by number of common locations, 
in Figure~\ref{fig:common_p_auc} .
We observe that the inference performance increases monotically with the number of common locations between two users. 
However, even when two users share no common locations,
our inference attack can still predict social links with fair performance (AUC around 0.7),
especially for New York and Los Angeles (AUC equal to 0.72).
This is essentially due to the fact that our inference attack 
takes mobility neighborhood into account. 
With a random walk method,
a user's mobility neighbors not only consist of locations he visits 
but also of users who visit the same locations as him and the locations these users visit. 
This enables to establish a connection between users sharing no common locations.
It is worth noting that our inference attack performs much better than baselines when there is no common location available between users. 
Indeed, the most effective baselines have AUCs close to 0.5 (equivalent to random guessing) when two users share no location in common.

\subsection{Attack with Geographical Grids}
\label{subsec:grid}
So far, our attack has been performed on fine-grained mobility data, i.e., check-ins at POIs.
However, in some cases, 
the adversary may not have access to mobility data with such fine-grained location information,
but only geo-coordinates (latitude and longitude).
In this subsection, we investigate whether our inference attack
is still effective in this situation.

To proceed, we partition the region covered by each city into geo-grids,
and assign a check-in into a grid if its geo-coordinates lie in the grid.
In our experiments, we have tried multiple granularity for partitioning,
including 0.0005\textdegree, 0.001\textdegree, 0.01\textdegree and 0.1\textdegree\ (similarly to the partitioning used in~\cite{CBCSHK10}). Results are presented in Figure~\ref{fig:grid_auc}.
At the finest granularity, i.e., 0.0005\textdegree\ (around 50m by 50m),
our inference attack achieves similar results as the case of POIs.\footnote{Note that there are already multiple POIs mapped to one single grid with 0.0005\textdegree, including POIs at the same latitude-longitude position but different height (e.g., in a building).}
We have AUC equal to 0.80 in Los Angeles, 0.79 in New York and 0.75 in London.
With geo-grids being coarser-grained,
the AUC values decrease monotonically.
However, even when the adversary only has the geo-coordinates at the granularity of 0.01\textdegree\
(around 1km by 1km),
our inference algorithm still performs quite well.
More interestingly, at the coarsest granularity, the AUC value is around 0.7 in Los Angeles,
while the results are much worse in New York and London.
This can be explained by different location densities in different cities.
Locations in Los Angeles are more uniformly distributed in the geo-space and distant from each other
than those in the other two cities.
In conclusion,
our attack is also effective when fine-grained location information
is not available, which further demonstrates the generality of our approach.

\section{Countermeasures}
\label{sec:defense}

In this section, we present three obfuscation mechanisms
for enhancing users' social relationship privacy 
while preserving the check-in dataset's utility as much as possible.
The three mechanisms, namely hiding, replacement and generalization,
are based on well-founded obfuscation schemes
proposed by the research community~\cite{STBH11,BHMSH15} 
for protecting users' location privacy.
We extend them here to protect users' social link privacy.
Since these defense mechanisms 
are not specific to a certain inference attack,
we evaluate them not only on our attack
but also on baseline models introduced in Section~\ref{sec:evalua}.

We first describe the utility metric considered in our defense,
then present the obfuscation mechanisms in detail and, finally, we experimentally study the performance of our defense.

\subsection{Utility Metric}
One approach to quantify utility is to
consider the global properties of the obfuscated dataset,
such as the check-in distribution over all locations in each city.
However, metrics of this kind neglect the individual check-in behavior, and could lead to obfuscated datasets becoming useless for a handful of applications, such as location recommendation~\cite{GTHL13,PZ17b}.
For keeping as much user utility as possible, we design a metric
which aims at measuring to what extent each user's check-in distribution is preserved.

We first denote a user $\user$'s check-in distribution in the original dataset as
$P^{o}_{\user}(\Visitloc)$ where $\Visitloc$ 
is the random variable to represent locations a user has visited.
Formally, 
\begin{equation}
P^{o}_{\user}(\Visitloc=\loc)=
  \begin{cases}
  \frac{\vert\uci(\user,\loc)\vert}{\vert\uci(\user)\vert} & \mbox{ if } \loc\in \uloc(\user),\\
  0 & \mbox { otherwise.}\\
  \end{cases}
\end{equation}
Accordingly, $\user$'s check-in distribution in the dataset 
obfuscated by a certain defense mechanism $\obf$
is defined as $P^\obf_\user(\Visitloc)$
and $P^\obf_\user(\Visitloc=\loc) = \frac{\vert\uci^\obf(\user,\loc)\vert}{\vert\uci^\obf(\user)\vert}$
for $\loc \in \uloc^\obf(\user)$.
Here,  $\uci^\obf(\user, \loc)$, $\uci^\obf(\user)$ and $\uloc^\obf(\user)$
denote $\user$'s check-ins at $\loc$, $\user$'s check-ins 
and the set of unique locations he has visited in the obfuscated dataset, respectively.
Then, $\user$'s utility loss is defined as the statistical distance 
between $P^{o}_\user(\Visitloc)$ and $P^\obf_\user(\Visitloc)$.
In this work, we adopt Jensen-Shannon divergence 
as the statistical distance.
Formally, $\user$'s \emph{utility loss} is defined as
\begin{equation}
\begin{aligned}
\utilityloss^\obf(\user) = \sum_{\loc\in \Loc}
P^{o}_\user(\Visitloc=\loc)\log_2\frac{P^{o}_\user(\Visitloc=\loc)}{M_\user(\Visitloc=\loc)} + \\
P^\obf_\user(\Visitloc=\loc)\log_2\frac{P^\obf_\user(\Visitloc=\loc)}{M_\user(\Visitloc=\loc)},
\end{aligned}
\end{equation}
where $M_\user(\Visitloc=\loc) = \frac{P^{o}_\user(\Visitloc=\loc)+P^\obf_\user(\Visitloc=\loc)}{2}$.
We use Jensen-Shannon divergence 
since it satisfies the symmetry property of a distance metric (contrary to the Kullback-Leibler divergence),
and has been used in previous works such as~\cite{MPS13}.
Moreover, Jensen-Shannon divergence 
lies in the range between 0 and 1 which allows us to easily define utility from the Jensen-Shannon divergence as follows:
\begin{equation}
\utility^\obf(\user) = 1-\utilityloss^\obf(\user).
\end{equation}
In the end, the utility of the whole dataset after applying $\obf$ 
is defined as the average utility loss over all users
\begin{equation}
\Utility^\obf = \sum_{\user\in\User}\frac{\utility^\obf(\user)}{\vert\User \vert}.
\end{equation}

\subsection{Obfuscation Mechanisms}
\label{subsec:obfuscate}
We now introduce the three obfuscation mechanisms for protecting social link privacy.

\smallskip
\noindent\textbf{Hiding.}
This mechanism simply removes a certain proportion of check-ins in the original dataset.
The check-ins to be removed are randomly sampled
and the remaining check-ins are used to calculate the utility 
following the previous definition.

\smallskip
\noindent\textbf{Replacement.}
This mechanism replaces a certain proportion of check-ins' locations with other locations
to mislead the adversary.
A location in a certain check-in
can be replaced by any location in the dataset.
In order to retain as much utility as possible, we adopt the random walk approach proposed by Mittal et al.~\cite{MPS13}
to find locations close to the original ones from a social mobility point of view.
For each check-in $\langle\user, t, \loc\rangle$ chosen to be replaced, 
we perform a random walk from $\user$ on the bipartite graph $\biG$
and replace the location of the check-in with the last node in the random walk trace.
Since $\biG$ is bipartite, the length of the random walk trace, another hyperparameter, needs to be odd
such that the random walk stops at a location (not at a user).
We empirically study how its length affects the performance of replacement with respect to inference performance and utility in the evaluation subsection.

It is worth noting that random walk used here
has a different purpose from the random walk used 
in the first stage of our inference attack (Section~\ref{sec:attack}).
The latter aims to reorganize $\biG$ into random walk traces for skip-gram
to learn each user's mobility features,
while the former utilizes the graph structure to find close locations
in order to keep the utility of the obfuscated dataset.

\smallskip
\noindent\textbf{Generalization.}
As presented in Section~\ref{sec:evalua},
for each location, we have its category information (collected from Foursquare)
and geo-coordinates, i.e., latitude and longitude.
Our third defense mechanism aims at generalizing
both the semantic and geographical dimensions.

Foursquare organizes its location categories\footnote{\url{https://developer.foursquare.com/categorytree}} 
into a two-level tree structure:
9 high-level categories
and 427 low-level categories.\footnote{This number 
is based on the result given by Foursquare's API in January 2016}
Therefore, for semantic generalization,
we logically rely on the two-category levels provided by Foursquare.
For geographical generalization,
we partition check-ins into geographic grids of different granularity (as in Section~\ref{subsec:grid}).
Here, we also consider two-level generalization:
0.01\textdegree\ (around 1 km by 1 km) grids for low-level generalization, 
and 0.1\textdegree\ (around 10 km by 10 km) grids for high-level generalization.
We consider 0.01\textdegree\ as low-level generalization and not 0.001\textdegree\
since, as shown in Figure~\ref{fig:grid_auc}, the inference performance with 0.001\textdegree\ grids is almost as good as for the original attack.
As in~\cite{BHMSH15}, geographic and semantic generalizations are considered jointly,
which gives us four different combinations of generalization, 
denoted by \emph{lg-ls} (low-level geo-grid, low-level semantics),
\emph{lg-hs} (low-level geo-grid, high-level semantics), 
\emph{hg-ls}  (high-level geo-grid, low-level semantics)
and \emph{hg-hs}  (high-level geo-grid, high-level semantics).

Different from hiding and replacement,
the generalization mechanism will modify the original set of locations (IDs) in the dataset
by merging multiple locations belonging to the same generalized location together.
However, when the adversary obtains the generalized dataset,
he can use external knowledge 
to map the generalized locations back to the original ones, and thereby increase the inference performance or utility provided to the user, respectively.
For instance, MoMA and Bernarducci Meisel Gallery in New York
are generalized into the same location under \emph{lg-hs},
i.e., art and entertainment place at geographic coordinates (40.76\textdegree\ N, -73.97\textdegree\ W).
When a user shares a check-in at this generalized location,
the attacker or service provider is more confident that the check-in 
is at MoMA than at Bernarducci Meisel Gallery,
since the former is much more popular than the latter.

In order to get conservative privacy guarantees for the generalization mechanism, 
we assume the adversary and service provider to be equipped with such external knowledge.
Practically, we construct the adversary's background knowledge by collecting each location's total number of check-ins from Foursquare's API (independently from the Instagram data).
For each check-in shared at a generalized location,
we sample a location that is included in this generalized location
as the check-in's original location
with a sampling rate equal to the proportion of check-ins
at this original location in the generalized location area.
\footnote{We do not consider external knowledge in Section~\ref{subsec:grid}
since we want to evaluate the performance of our attack.
In that case, a simple adversary is a reasonable choice.
On the other hand, for evaluating the generalization mechanism and get safe privacy guarantees,
it is necessary to consider a stronger adversary with external knowledge.}

\subsection{Defense Evaluation}

We evaluate all the three obfuscation mechanisms 
against our inference attack as well as baseline models.
Both hiding and replacement mechanisms 
involve randomly obfuscating 
a certain proportion of check-ins in the original dataset.
In our experiments, we choose to hide or replace from 10\% to 90\% check-ins in incremental steps of 10\%.
For presentation purposes, we only depict the results for New York,
results for Los Angeles and London following a similar trend and being presented in Appendix~\ref{appendix:defense}.

Figure~\ref{fig:defense_edge_weight_modifyrw}
presents our inference attack's performance against hiding and replacement.
We observe that replacement is more effective than hiding 
on decreasing our inference attack's performance 
when the proportion of obfuscated check-ins is fixed.
For instance,
when obfuscating 30\% of check-ins, 
replacement decreases our attack's AUC by 7\%
while hiding only decreases it by 3\%. Moreover, in order to degrade the inference performance sufficiently to make a poor prediction (AUC < 0.7), we need to hide 80\% of the check-ins or replace 50\% of them.
This is due to the fact that 
the replacement mechanism introduces more noise to the original dataset than randomly hiding check-ins,
which will result in skip-gram learning less informative features for each user. 
However, as hiding does not cause significant changes to a user's mobility distribution,
it preserves more utility
than replacement for a fixel level of obfuscation (Figure~\ref{fig:utility_edge_weight_modifyrw}). 
This demonstrates that there exists a tension between privacy and utility in social link privacy protection, and that there is no free lunch in such a setting.

We empirically evaluate the impact of the number of steps considered in the random walk for the replacement mechanism.
Our experiments show that increasing the steps from 5 to 15 
decreases attack performance quite significantly (Figure~\ref{fig:defense_edge_weight_modifyrw}), but that further step increase does not provide much more privacy to the users (as the AUC value then saturates for all obfuscation proportions).
The same decreasing behavior holds for utility, but the difference is much smaller between 5 steps and 15, 25 and 35 steps than for the AUC value decrease.
By further taking into account the computational time
(bigger walk steps leads to longer execution time),
we believe that 15 provides the best trade-off between privacy, utility, and efficiency
for the replacement mechanism.
Figure~\ref{fig:baseline_edge_weight_modifyrw} further shows AUC
for hiding and replacement against the three best performing baseline models, i.e.,
$\mathtt{w\_common\_p}$, $\mathtt{common\_p}$ and $\mathtt{overlap\_p}$.
As for our attack, replacement is more effective than hiding on decreasing the AUC of the baselines for all proportions of obfuscation except 90\%.

\begin{figure*}[!t]
\begin{subfigure}{0.69\columnwidth}
\includegraphics[width=\columnwidth]{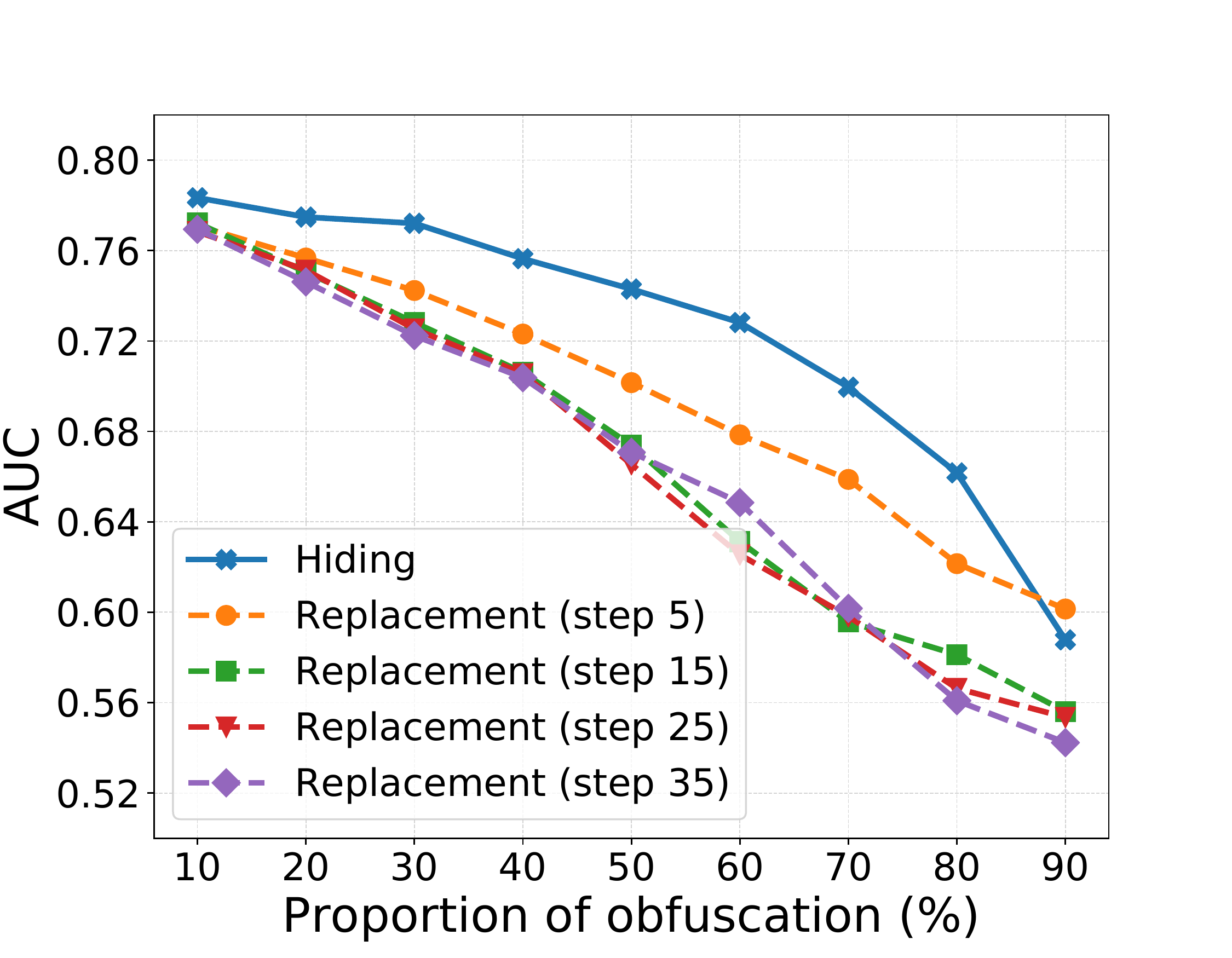}
\caption{}
\label{fig:defense_edge_weight_modifyrw}
\end{subfigure}
\begin{subfigure}{0.69\columnwidth}
\includegraphics[width=\columnwidth]{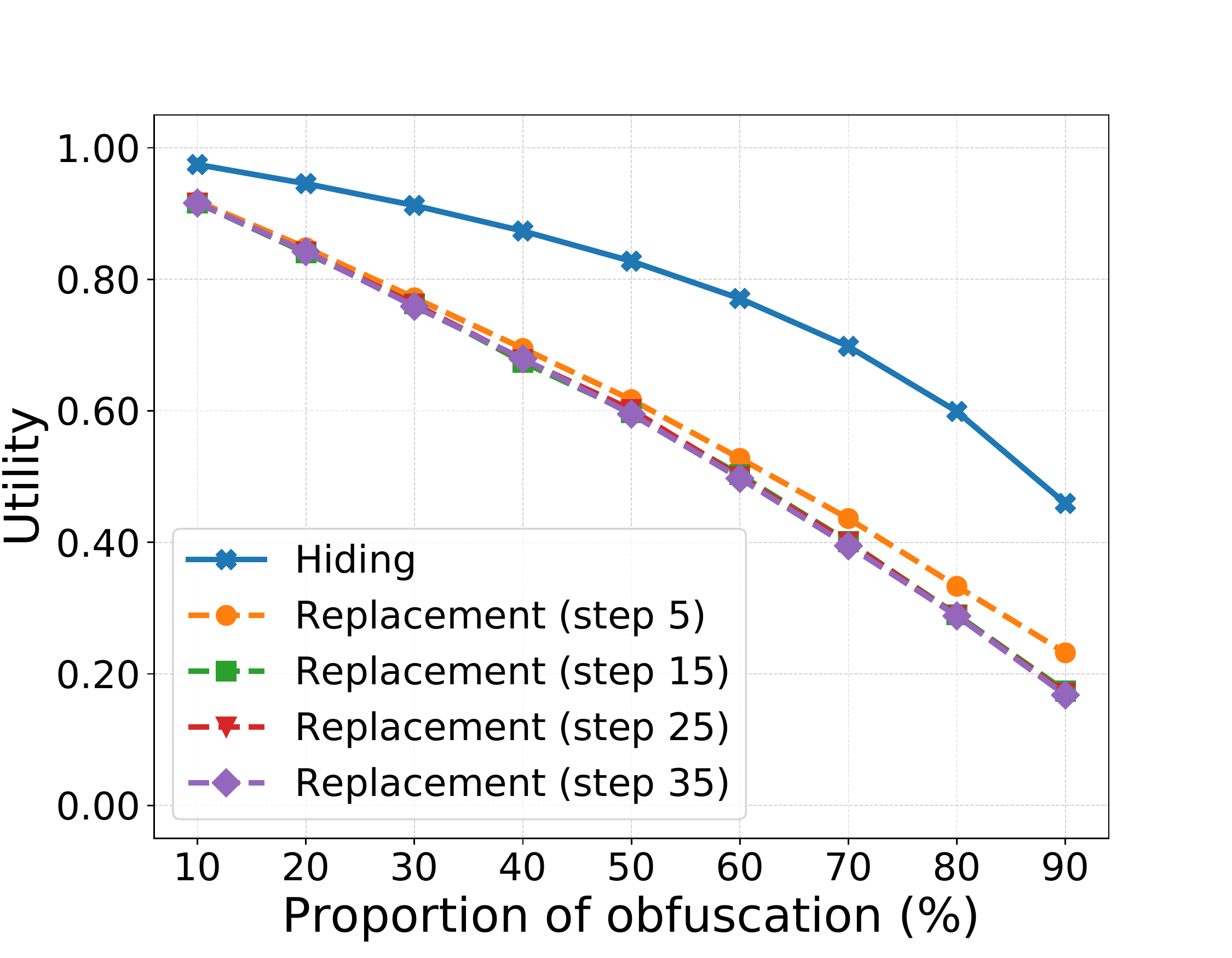}
\caption{}
\label{fig:utility_edge_weight_modifyrw}
\end{subfigure}
\begin{subfigure}{0.69\columnwidth}
\includegraphics[width=\columnwidth]{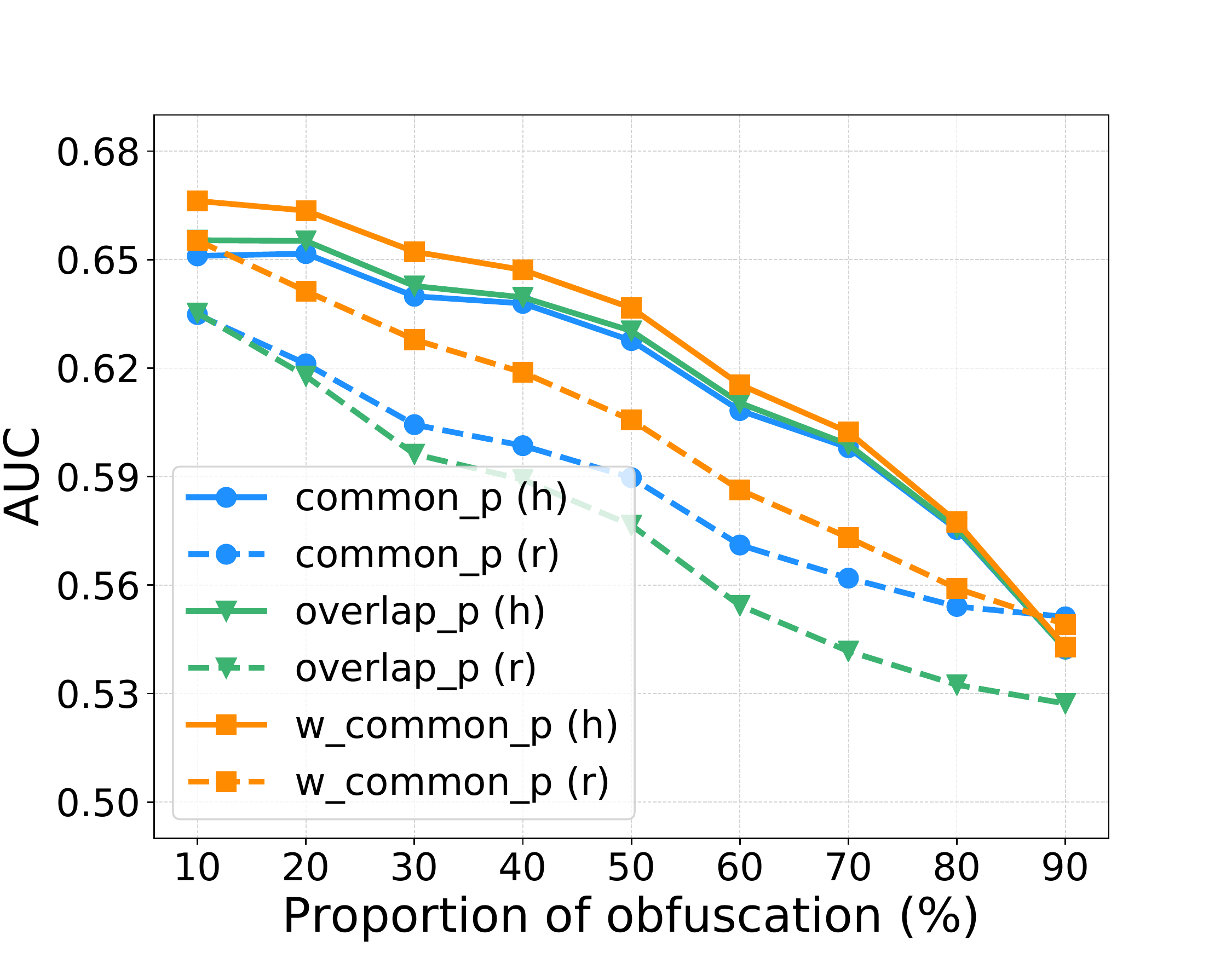}
\caption{}
\label{fig:baseline_edge_weight_modifyrw}
\end{subfigure}
\caption{Hiding vs.\ replacement with respect to the proportion of obfuscation: (a) inference performance (AUC) of our attack,
(b) utility, and (c) inference performance (AUC) of baseline models in New York.
The length of random walk steps in replacement is set to 15 for baseline models,
h represents hiding and r stands for replacement.}
\label{fig:hide_replace}
\end{figure*} 

Table~\ref{table:gene} presents the AUC values
and utility of the generalization mechanism (for our attack and the three best baselines).
First, we observe that higher-level generalization leads to the worst inference performance, thus best privacy provision, as expected. However, we also notice that utility is decreased a lot with this countermeasure, down to 0.06 for maximal generalization.
Interestingly, the lowest-level generalization, i.e., \emph{lg-ls},
is not very helpful for social link privacy (AUC = 0.77 compared to AUC = 0.79 without countermeasure) for a utility decrease that is still substantial.
This indicates that generalization does not provide an optimal balance between utility and privacy. This is essentially due to the fact that the external knowledge (about location popularity)
helps the adversary improve his inference attack in presence of this countermeasure.

Second, \emph{lg-hs} provides a better inference performance and utility than \emph{hg-ls},
which means that getting more precise geo-coordinates is more informative about social relationships than having more precise semantic information. Nevertheless, by comparing results from Figure~\ref{fig:grid_auc} in Section~\ref{subsec:grid} to those reported here, we clearly observe that semantic information brings a lot of information to the adversary (as shown in~\cite{AHHH16} for location inference). Indeed, we notice that the AUC here with \emph{hg-hs} is equal to 0.67 whereas it is equal to around 0.6 in Figure~\ref{fig:grid_auc} with similiar geographic information but no semantics. Hence, we see that even high-level semantic information brings sufficient knowledge to increase the attack's AUC by 12\%. Lower-level semantic data increases it by 22\% to 0.73.

We further calculate the adversary's recovery rate, 
i.e., the proportion of original check-ins that are recovered.
The results are presented in Table~\ref{table:gene} too.
As we can see, when the generalization level is \emph{lg-ls},
the adversary is able to recover 52\% of the original location IDs.
Given that we only use a very simple recovery algorithm based on the global locations' distribution,
this confirms that generalization is not enough
to protect location and social link privacy against adversaries with external knowledge.
Moreover, \emph{lg-hs} has a higher location recovery rate than \emph{hg-ls} (23\% vs.\ 14\%),
which also explains why the attacker achieves a higher AUC in \emph{lg-hs} than in \emph{hg-ls}.

\begin{table}[!h]
\centering
\caption{Inference performance and utility for generalization in New York.}
\label{table:gene}
\begin{tabular}{| c | cc | cc | cc |}
\hline
& \multicolumn{2}{|c|}{AUC} & \multicolumn{2}{|c|}{Utility} & \multicolumn{2}{|c|}{Recovery rate}\\ 
\hline
&\emph{ls}&\emph{hs} &\emph{ls}&\emph{hs} &\emph{ls}&\emph{hs}\\
\hline
\emph{lg}&0.77&0.75 & 0.57&0.30 & 52\% & 23\% \\
\emph{hg}& 0.73 & 0.67 &0.20&0.06 & 14\% & 2\% \\
\hline
\hline
& \multicolumn{2}{|c|}{$\mathtt{w\_common\_p}$} & \multicolumn{2}{|c|}{$\mathtt{overlap\_p}$}
& \multicolumn{2}{|c|}{$\mathtt{common\_p}$}\\
\hline
 &\emph{ls}&\emph{hs}  & \emph{ls}&\emph{hs} & \emph{ls}&\emph{hs}\\
\hline
\emph{lg} & 0.65&0.63 & 0.65 & 0.63 & 0.65 & 0.64\\
\emph{hg} & 0.61&0.58 & 0.60 & 0.57 & 0.62 & 0.58\\
\hline
\end{tabular}
\end{table}

\begin{figure}[!t]
\centering
\includegraphics[width=0.85\columnwidth]{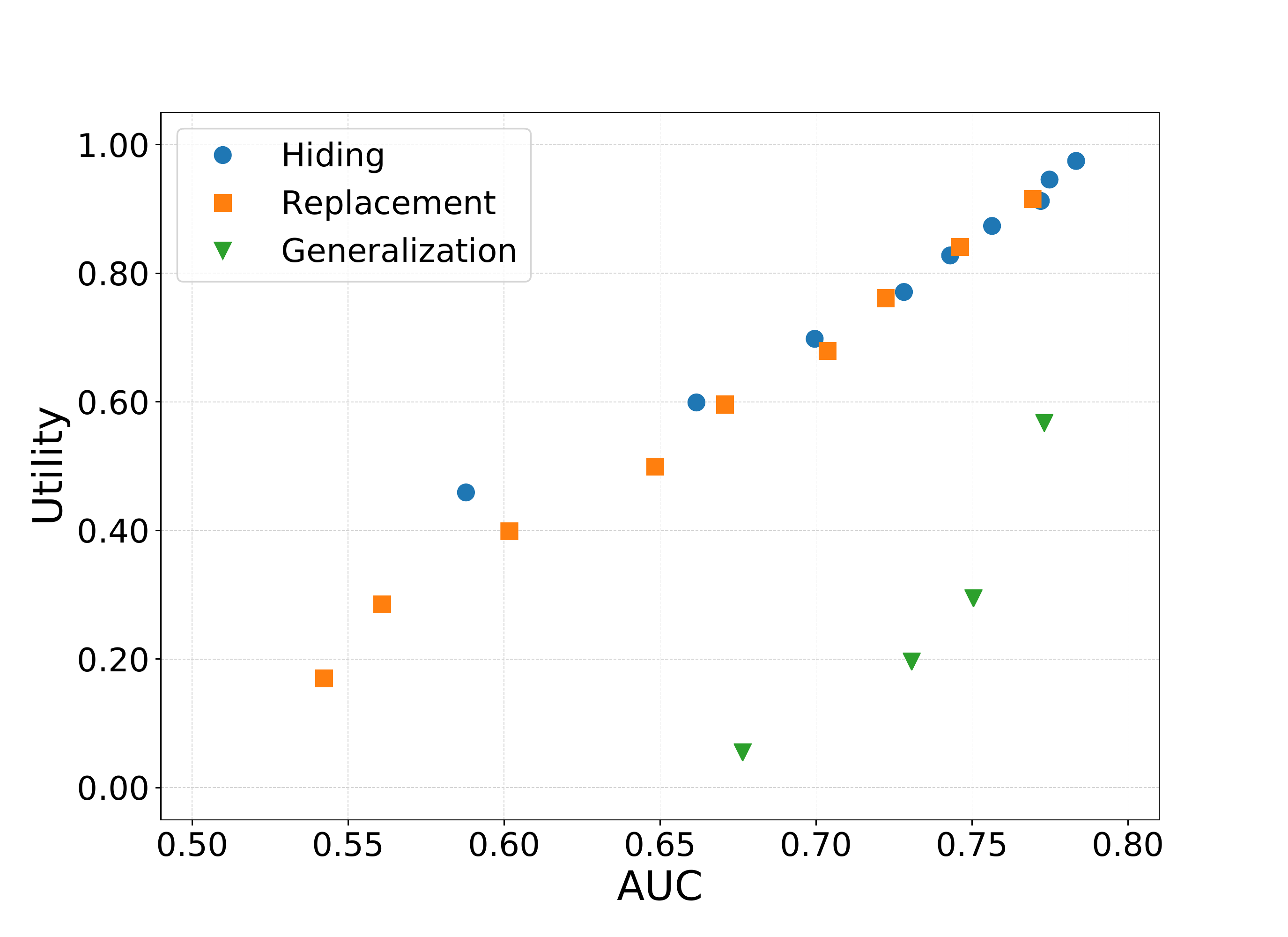}
\caption{AUC vs.\ utility for three obfuscation mechanisms.}
\label{fig:auc_utility}
\end{figure}

When comparing the three obfuscation mechanisms by fixing the AUC value (with our inference attack),
hiding and replacement achieve a comparable performance in general,
and they both outperform generalization (Figure~\ref{fig:auc_utility}).
For instance, if we want to achieve a utility of at least 0.6, then the AUC values
of hiding and replacement are very close to each other, of 0.66 and 0.67, respectively.
However, we observe that, for a similar AUC value, utility drops to 0.06 with the generalization mechanism.
From Figure~\ref{fig:auc_utility},
it seems that hiding performs better than replacement.
But we should also notice that 
replacement can decrease inference attack's performance, thus improve privacy,
to a larger extent than hiding:
when obfuscating 90\% check-ins,
replacement decreases our attack's AUC to 0.54,
while hiding only leads to a minimal AUC of 0.59.

\section{Related Work}
\label{sec:related}

With the increasing usage of portable devices,
a large amount of mobility data has become available. On the one hand, this represents an unprecedented chance 
to study the interaction between human mobility and social networks. On the other hand, it raises new concerns towards privacy.
In the following, we separate the most related literature into two main research topics.
The first line of research concentrates on inferring hidden location information from social data while the second line focuses on leveraging mobility data to infer social relationships.

Backstrom et al.~\cite{BSM10} develop a maximal likelihood estimator
to predict a user's undisclosed home location with his friends' data.
Experiments on a large Facebook dataset 
show that their model outperforms traditional IP-based approaches significantly.
Following this work, the authors of~\cite{MCC13} have incorporated 
fine-grained social relation information into their home location prediction model.
Evaluation on a Twitter dataset has demonstrated that 
social features such as number of followers
indeed increase the prediction performance.
Cho et al.~\cite{CML11} have observed on a Gowalla dataset
that a user's mobility is centered around two states: home and work.
They develop a Gaussian mixture model to learn the two hidden states
and further incorporate friendship influence.
Extensive experiments demonstrate the effectiveness of their approach.
Recently,
Olteanu et al.~\cite{OHSHH17} have shown how co-location information about OSN users (e.g., via location check-ins with two or more users) can be used by an attacker to degrade these users' location privacy. They provide an analytical framework based on Bayesian networks to formally quantify the effects of co-location data on location privacy and also consider the impact of some obfuscation mechanisms. 
Other interesting works in this direction include~\cite{SKB12,J13,JFMXR15,PZ15,WL16}.

The second line of research tackles the dual problem,
i.e., using mobility data to infer the underlying social relations.
Our inference attack and the baseline models~\cite{SNM11,PSL13,WLL14} we compare it to
fall into this topic.
Eagle et al.~\cite{EPL09} have first shown that there exist correlations between people's co-occurrences
and their social connections by conducting a study based on mobile phone records.
Crandall et al.~\cite{CBCSHK10} go one step beyond by relying on a Bayesian model 
to show that the friendship probability of two users with joint mobility behavior
is 5,000 times higher than those without joint behavior.
These results shed light on the social relation privacy threat carried by mobility data.
However, the model they use
makes an over-simplified assumption that each user only has one friend.

Scellato et al.~\cite{SNM11} tried to get closer to a realisitic setting by
proposing 15 novel machine learning features.
Among the 15 features, 
4 of them follow a classical link prediction setting~\cite{LK07} 
by relying on some existing social network structure.
In our work, we assume that our adversary has no knowledge of any existing social links.
Besides, we evaluate all the other 11 features as part of the baseline models.
Moreover, their evaluation is conducted on some predefined inference spaces
such as two users need to share common friends or common locations. 
In our experiments, we do not impose any constraint on the mobility profiles of users, 
and thus make a more realistic evaluation of these baseline models and our inference attack.

Pham et al.~\cite{PSL13} propose two features for social link inference,  
i.e., $\mathtt{diversity}$ and $\mathtt{w\_frequency}$.
The former concentrates on the diversity of two users' joint check-in behaviors
and the latter reflects the popularity of two users' common locations.
Both $\mathtt{diversity}$ and $\mathtt{w\_frequency}$ are based on entropy measures.
The authors of~\cite{WLL14} propose three mobility factors, 
namely personal, global and temporal,
Among them, 
the global factor is the same as $\mathtt{w\_frequency}$ in~\cite{PSL13},
while the personal factor ($\mathtt{personal}$)
follows the intuition that two users are more likely to know each other
if they meet at locations they do not visit frequently.

Different from~\cite{SNM11},
both~\cite{PSL13} and~\cite{WLL14}
consider two users' meeting events (visiting the same location at roughly the same time) 
instead of common locations.
However, meeting events are really rare even in our large dataset,
meaning that the methods in~\cite{PSL13} and~\cite{WLL14}
can only apply to a small set of users.
Even when we concentrate on users with meeting events,
features in~\cite{PSL13}  and~\cite{WLL14}
do not achieve any performance gain compared to the case of common locations,
especially for $\mathtt{personal}$ in~\cite{WLL14}, where
the performance even worsens.
Therefore, we decide to use common locations as in~\cite{SNM11} 
instead of meeting events
to evaluate the baselines in~\cite{PSL13} and~\cite{WLL14}.\footnote{
Following the same reason,
we do not implement the temporal factor in~\cite{WLL14} as one baseline model.}
As shown in Section~\ref{sec:evalua},
our inference attack significantly outperforms these baselines,
which demonstrates the effectiveness and relevance of our approach.


\section{Conclusion}
\label{sec:conclusion}

Mobility data are nowadays largely available to a wide range of service providers. This raises many privacy issues, especially when such providers' data ends up into the hands of intelligence agencies. This paper aims at evaluating, with a principled approach, the impact on social link privacy of this wide availability of location data. To this endeavor, we propose a new generic method for inferring social links without imposing any prior condition on users' mobility patterns. Furthermore, we design countermeasures for mitigating the extent of the privacy threat towards social relationships.

The empirical evaluation of our inference attack demonstrates that our principled approach outperforms previously proposed inference algorithms by up to 20\% on a large-scale dataset, with an area under the ROC curve of around 0.8. Our results further show that our attack provides fair prediction results (AUC equal to 0.71 or 0.75 depending on the targeted city) even when the number of available location points per user is small (down to 5). Moreover, our attack is quite robust to a low number of common locations between two users. For two cities, it even provides fair prediction performance (AUC around 0.72) when two users share no location at all in common. Finally, we observe that our attack performs also well with geographic grids of size up to 1-by-1 km instead of exact semantic and geographic location data.

In order to counter the presented attack against social link privacy, we propose and evaluate three well-established privacy-preserving techniques: hiding, replacement and generalization. Our empirical results demonstrate that, in order to degrade the inference performance sufficiently to make a poor prediction (AUC smaller than 0.7), we need to hide 80\% of the location points or replace 50\% of them. However, we notice also that replacement decreases utility more than hiding, which shows that there is no free lunch in such a privacy setting. Furthermore, we notice that the generalization mechanism provides a much poorer privacy-utility trade-off than the hiding and replacement techniques. Finally, by comparing our defense and attack results, we observe that the semantic dimension of locations can have substantial positive effect on the social link inference when geographic information is obfuscated with generalization.

\begin{acks}
This work was partially supported by the German Federal Ministry of Education and
Research (BMBF) through funding for the Center for IT-Security,
Privacy and Accountability (CISPA) (FKZ: 16KIS0656).
Part of this work was carried out while Mathias Humbert was with CISPA, Saarland University.
The authors would like to thank Rose Hoberman, Jonas Schneider and Kathrin Grosse 
for their valuable comments on the submitted manuscript. 
\end{acks}
\bibliographystyle{ACM-Reference-Format}
\bibliography{walk2friends} 
\appendix

\section{Appendix}

\subsection{Pairwise Similarity Measurements}
\label{appendix:pairwise}
We present the formal definitions of the 7 pairwise similarity measurements
used in our evaluation.

\noindent\textbf{Cosine similarity.}
\[
s(\feature{u}, \feature{u'}) =
\frac{\feature{u}\cdot\feature{u'}}{\vert \vert\feature{u}\vert\vert_2\ \vert \vert\feature{u'}\vert\vert_2}
\]

\noindent\textbf{Euclidean distance.}
\[
s(\feature{u}, \feature{u'}) =
\vert \vert\feature{u}-\feature{u'}\vert\vert_2 
\]

\noindent\textbf{Correlation coefficient.}
\[
s(\feature{u}, \feature{u'}) =
\frac{(\feature{u}-\overline{\feature{u}})\cdot(\feature{u'}-\overline{\feature{u'}})}{\vert \vert\feature{u}-\overline{\feature{u}}\vert\vert_2\ \vert \vert\feature{u'}-\overline{\feature{u'}}\vert\vert_2}
\]
Here, $\overline{\feature{u}}$ represents the mean value of $\feature{u}$.

\noindent\textbf{Chebyshev distance.}
\[
s(\feature{u}, \feature{u'}) = \max_{i= 1}^{d}\vert \feature{u}_i-\feature{u'}_i\vert
\]
Here, $\feature{u}_i$ represents the $i$th element in $\feature{u}$.

\noindent\textbf{Bray-Curtis distance.}
\[
s(\feature{u}, \feature{u'}) = \frac{\sum_{i= 1}^{d}\vert \feature{u}_i-\feature{u'}_i\vert}{\sum_{i= 1}^{d}\vert \feature{u}_i+\feature{u'}_i\vert}
\]

\noindent\textbf{Canberra distance.}
\[
s(\feature{u}, \feature{u'}) = \sum_{i= 1}^{d}\frac{\vert \feature{u}_i-\feature{u'}_i\vert}{\vert \feature{u}_i\vert + \vert\feature{u'}_i\vert}
\]

\noindent\textbf{Manhattan distance.}
\[
s(\feature{u}, \feature{u'}) = \sum_{i= 1}^{d}\vert \feature{u}_i-\feature{u'}_i\vert
\]

\subsection{Defense Evaluation for Los Angeles and London}
\label{appendix:defense}

The defense evaluation results for Los Angeles and London are presented as the following.

\begin{figure*}[!t]
\begin{subfigure}{0.69\columnwidth}
\includegraphics[width=\columnwidth]{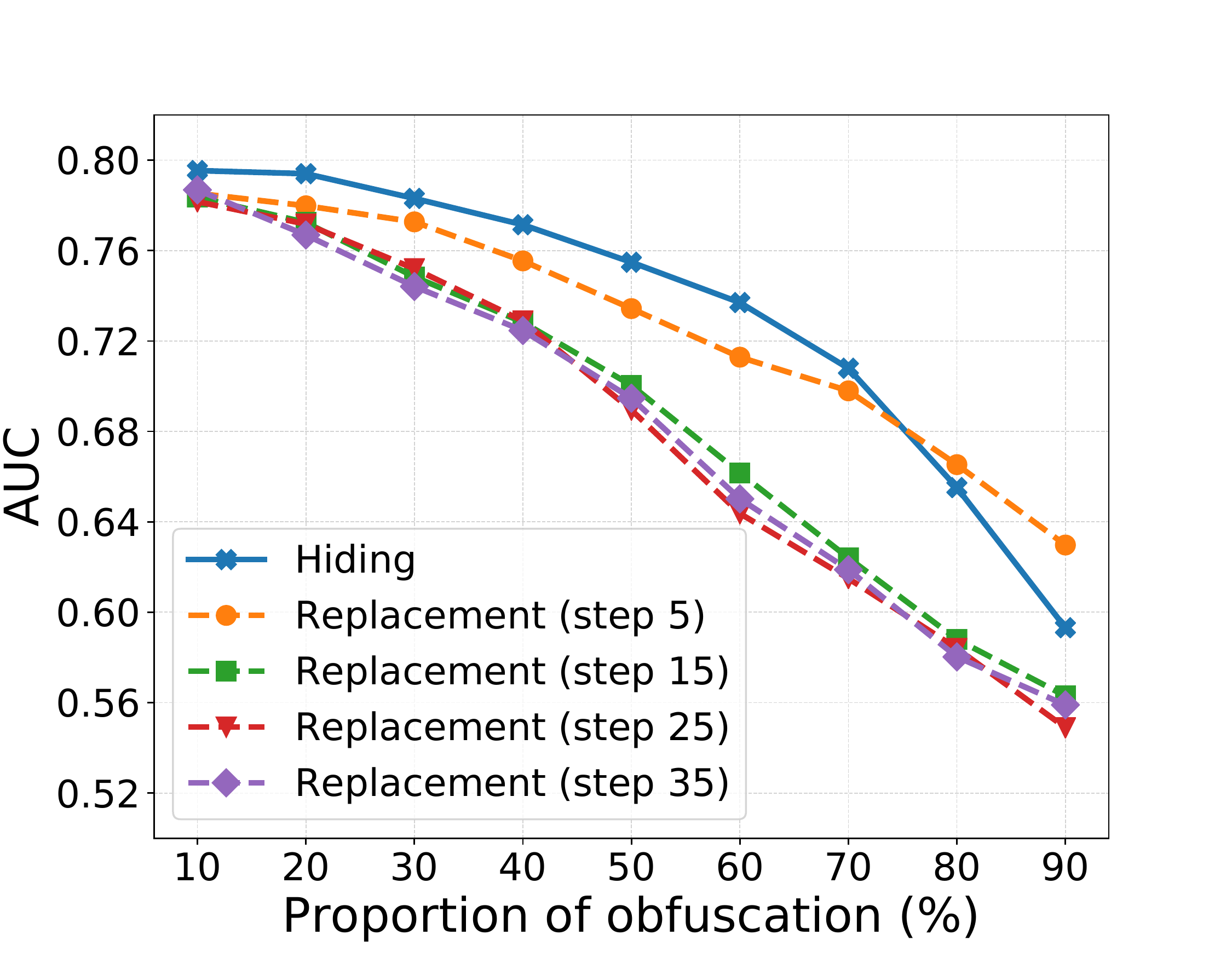}
\caption{}
\end{subfigure}
\begin{subfigure}{0.69\columnwidth}
\includegraphics[width=\columnwidth]{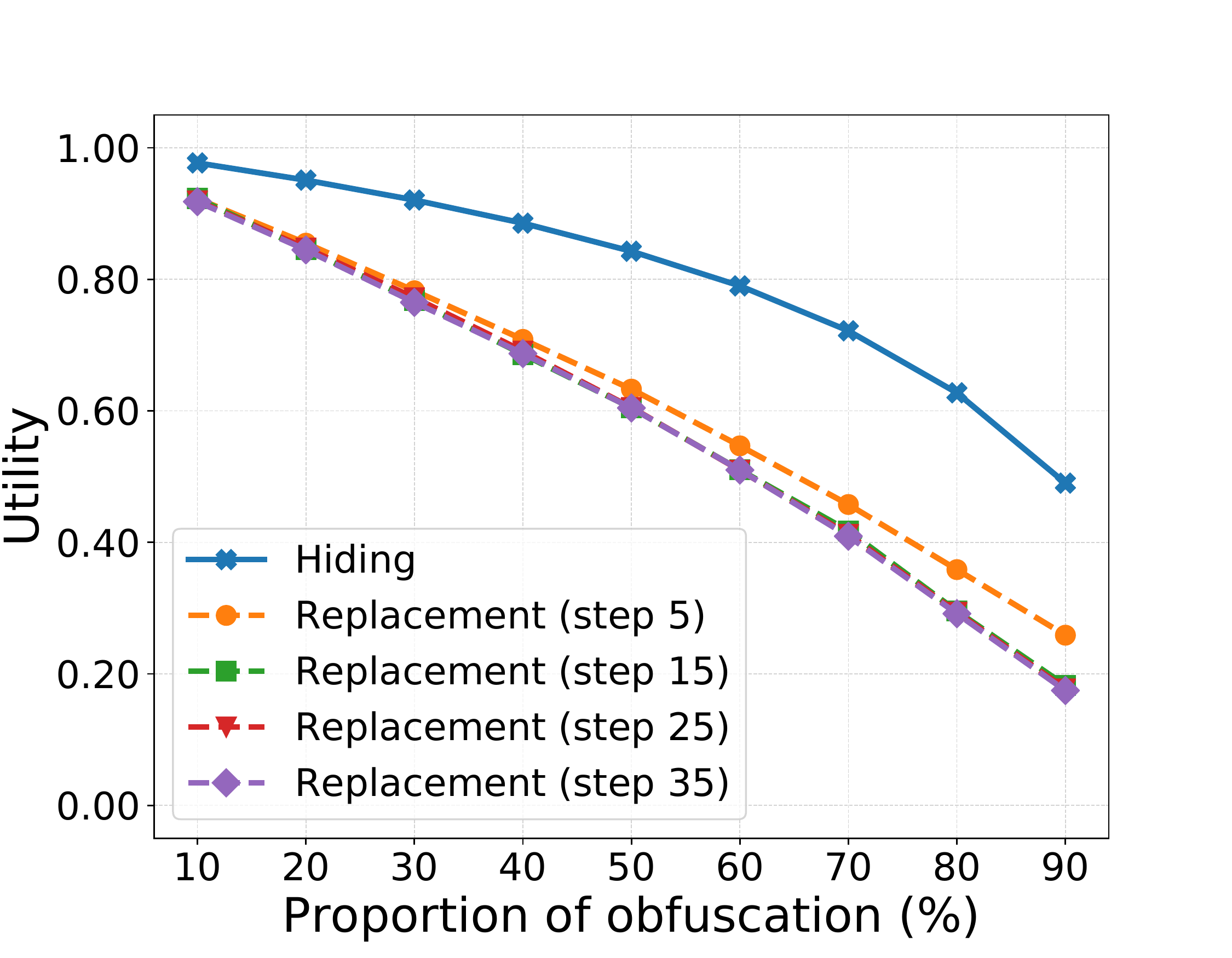}
\caption{}
\end{subfigure}
\begin{subfigure}{0.69\columnwidth}
\includegraphics[width=\columnwidth]{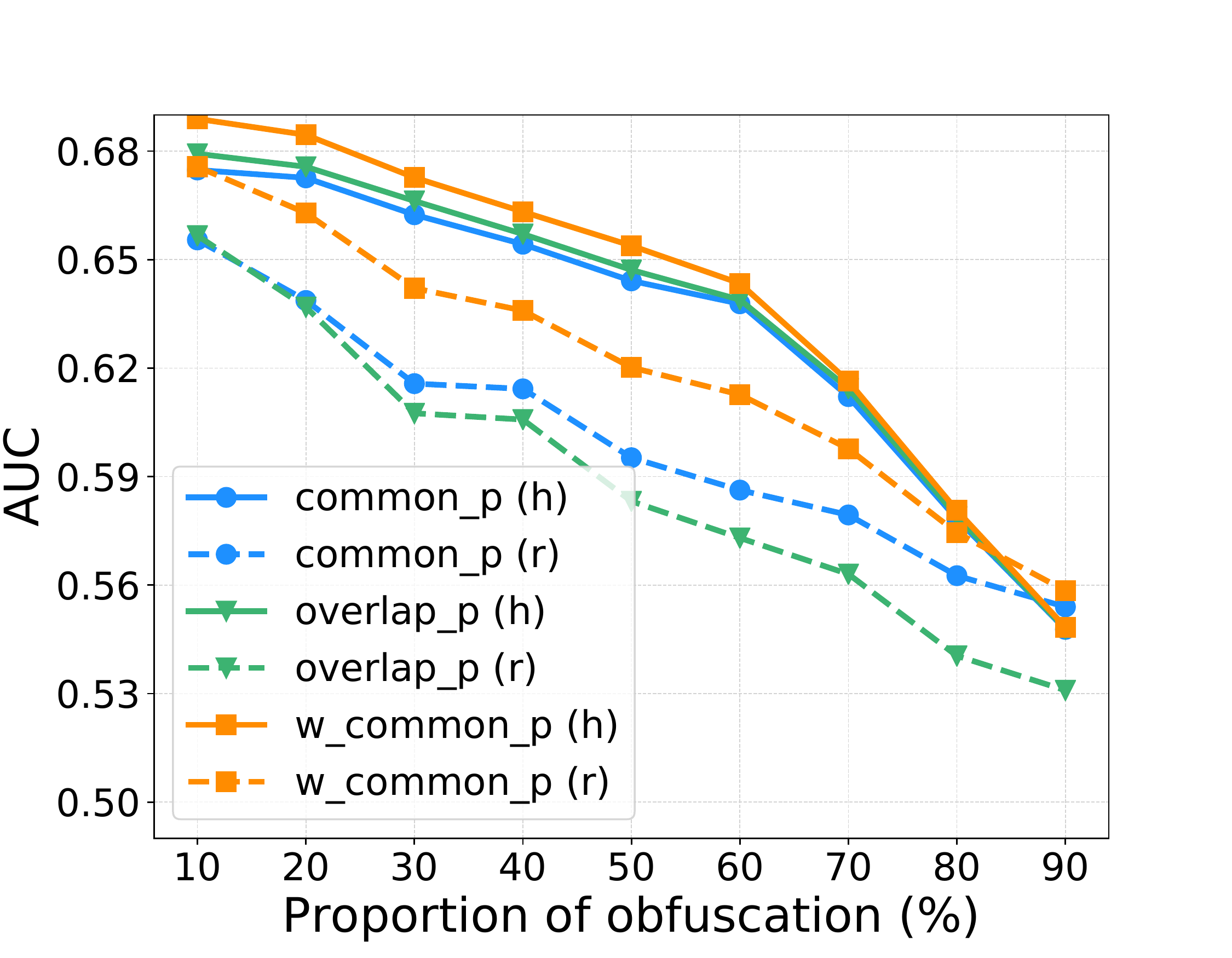}
\caption{}
\end{subfigure}
\caption{Hiding vs.\ replacement with respect to (a) inference performance on our attack,
(b) utility and (c) inference performance on baseline models in Los Angeles.
The length of random walk steps in replacement is 15 for baseline models,
h represents hiding and r represents replacement.}
\end{figure*} 

\begin{figure*}[!t]
\begin{subfigure}{0.69\columnwidth}
\includegraphics[width=\columnwidth]{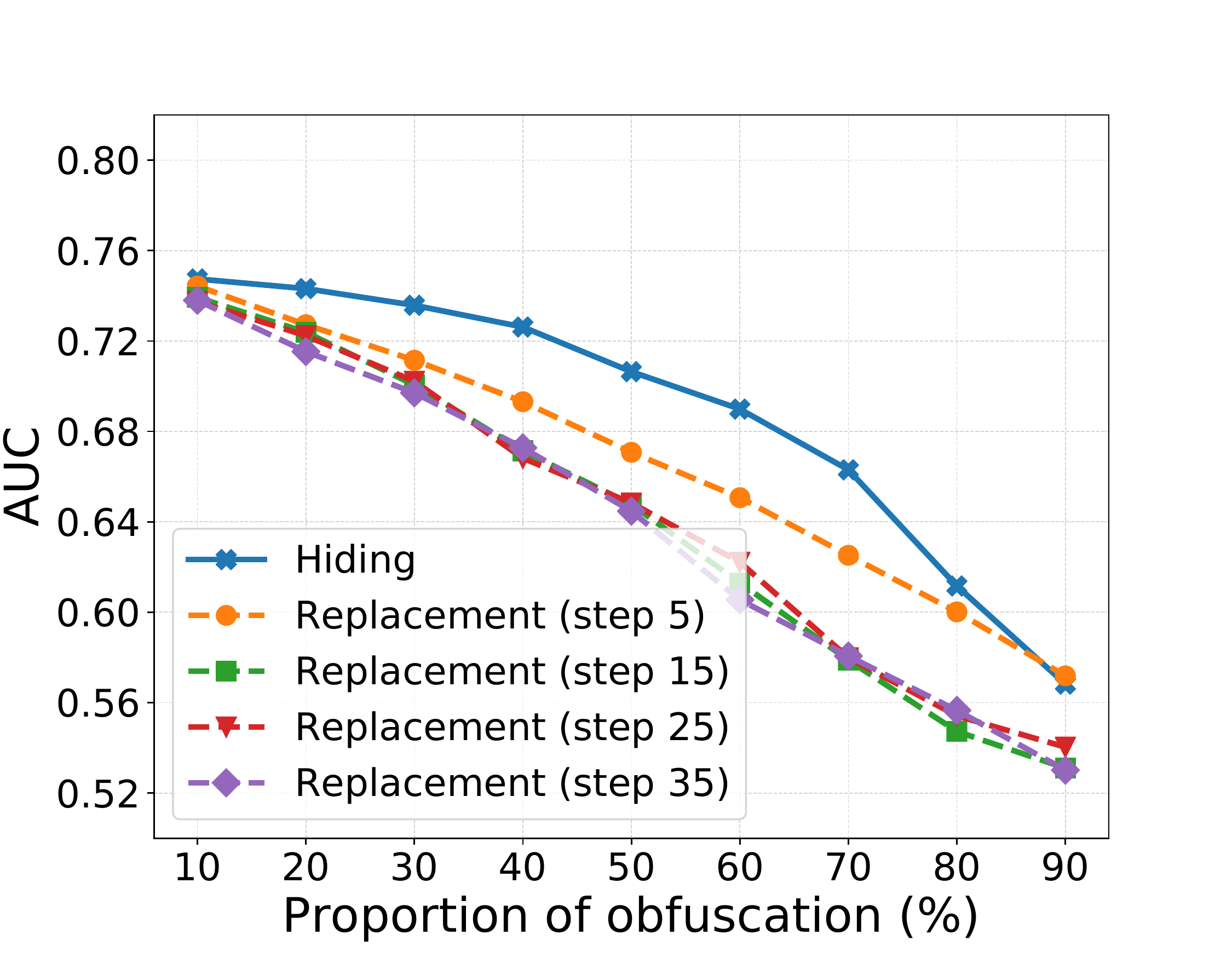}
\caption{}
\end{subfigure}
\begin{subfigure}{0.69\columnwidth}
\includegraphics[width=\columnwidth]{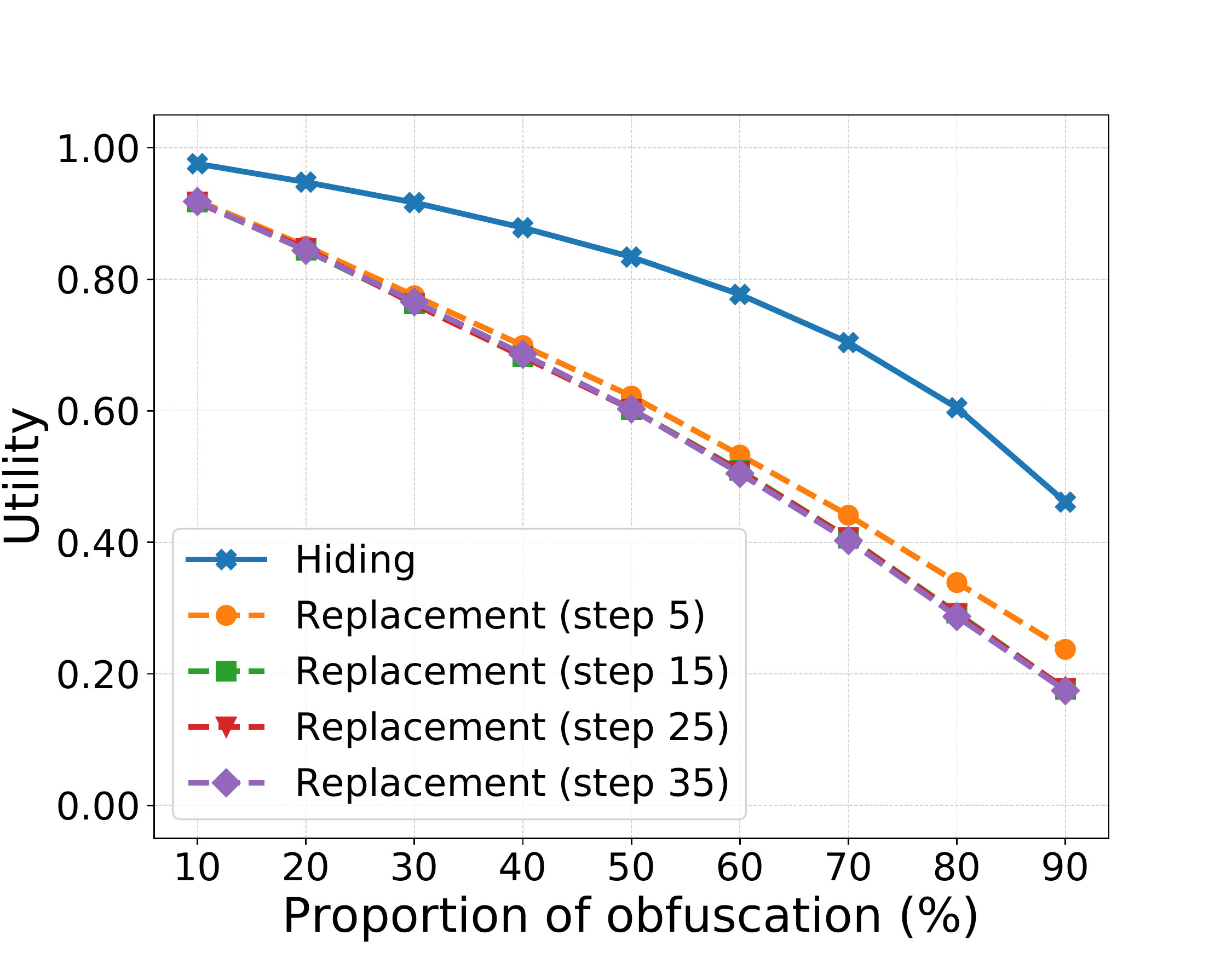}
\caption{}
\end{subfigure}
\begin{subfigure}{0.69\columnwidth}
\includegraphics[width=\columnwidth]{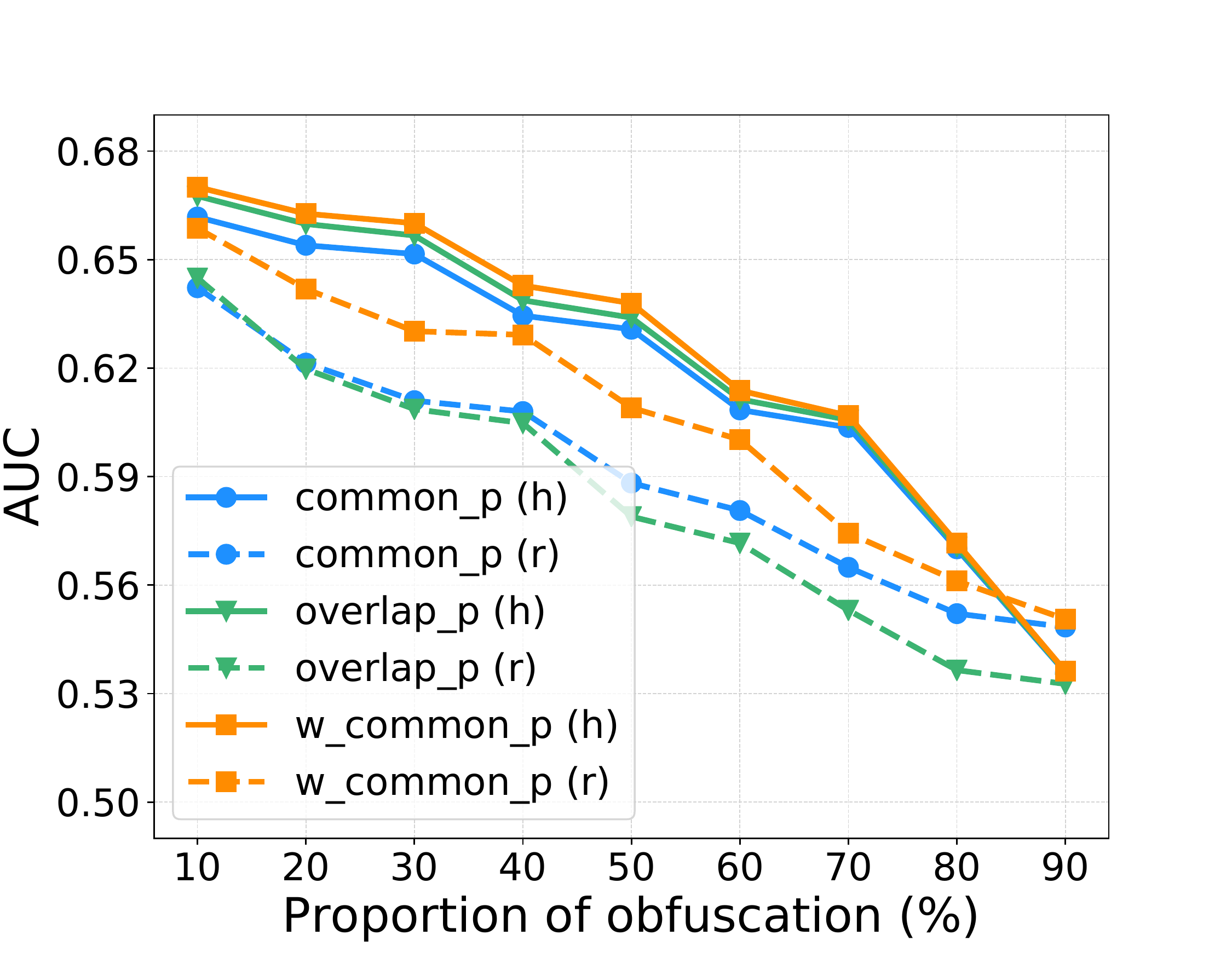}
\caption{}
\end{subfigure}
\caption{Hiding vs.\ replacement with respect to (a) inference performance on our attack,
(b) utility and (c) inference performance on baseline models in London.
The length of random walk steps in replacement is 15 for baseline models,
h represents hiding and r represents replacement.}
\end{figure*}

\begin{table}[h]
\centering
\caption{Inference performance and utility for generalization in Los Angeles.}
\begin{tabular}{| c | cc | cc | cc |}
\hline
& \multicolumn{2}{|c|}{AUC} & \multicolumn{2}{|c|}{Utility} & \multicolumn{2}{|c|}{Recovery rate}\\ 
\hline
&\emph{ls}&\emph{hs} &\emph{ls}&\emph{hs} &\emph{ls}&\emph{hs}\\
\hline
\emph{lg}&0.79&0.78 & 0.79&0.48 & 74\% & 40\% \\
\emph{hg}& 0.77 & 0.74 &0.37&0.13 & 29\% & 7\% \\
\hline
\hline
& \multicolumn{2}{|c|}{$\mathtt{w\_common\_p}$} & \multicolumn{2}{|c|}{$\mathtt{overlap\_p}$}
& \multicolumn{2}{|c|}{$\mathtt{common\_p}$}\\
\hline
 &\emph{ls}&\emph{hs} & \emph{ls}&\emph{hs} & \emph{ls}&\emph{hs}\\
\hline
\emph{lg} & 0.68&0.67 & 0.68 & 0.66 & 0.68 & 0.67\\
\emph{hg} & 0.66&0.63 & 0.66 & 0.63 & 0.66 & 0.64\\
\hline
\end{tabular}
\end{table}

\begin{table}[h]
\centering
\caption{Inference performance and utility for generalization in London.}
\begin{tabular}{| c | cc | cc | cc |}
\hline
& \multicolumn{2}{|c|}{AUC} & \multicolumn{2}{|c|}{Utility} & \multicolumn{2}{|c|}{Recovery rate}\\ 
\hline
&\emph{ls}&\emph{hs} &\emph{ls}&\emph{hs} &\emph{ls}&\emph{hs}\\
\hline
\emph{lg}&0.74&0.72 & 0.72&0.43 & 68\% & 36\% \\
\emph{hg}& 0.71 & 0.66 &0.28&0.08 & 21\% & 4\% \\
\hline
\hline
& \multicolumn{2}{|c|}{$\mathtt{w\_common\_p}$} & \multicolumn{2}{|c|}{$\mathtt{overlap\_p}$}
& \multicolumn{2}{|c|}{$\mathtt{common\_p}$}\\
\hline
 &\emph{ls}&\emph{hs} & \emph{ls}&\emph{hs} & \emph{ls}&\emph{hs}\\
\hline
\emph{lg} & 0.65&0.63 & 0.66 & 0.63 & 0.65 & 0.63\\
\emph{hg} & 0.63&0.59 & 0.62 & 0.58 & 0.62 & 0.59\\
\hline
\end{tabular}
\end{table}

\end{document}